\documentclass[aps,prc,preprint,showpacs,showkeys,byrevtex]{revtex4}
\usepackage{graphicx}
\usepackage{color}
\usepackage{dcolumn}

\newcommand{\bmath}[1]{\mbox{\boldmath${#1}$}}
\newcommand{\ri}[1]{{\bf r}_{#1}}

\newcommand{\ki}[1]{{\bf k}_{#1}}
\newcommand{\bp}{{\bf p}}
\newcommand{\qi}[1]{{\bf q}_{#1}}

\newcommand{\bsigma}{\bmath{\sigma}}

\newcommand{\bpi}{\bmath{\pi}}
\newcommand{\beps}{\bmath{\epsilon}}

\newcommand{\khat}{\bmath{\widehat{\bf k}}}
\newcommand{\phat}{\bmath{\widehat{\bf p}}}
\newcommand{\qhat}{\bmath{\widehat{\bf q}}}
\newcommand{\rhat}{\bmath{\widehat{\bf r}}}

\newcommand{\expup}[1]{e^{#1}}

\newcommand{\EG}{{\textrm{e.g.}}}
\newcommand{\IE}{{\textrm{i.e.}}}
\newcommand{\EA}{{\textit{et al.}}}

\date{\today}

\begin{document}
\title{Using chiral perturbation theory to extract the neutron-neutron 
scattering length from \boldmath$\pi^-d\to nn\gamma$}

\author{A. G{\aa}rdestig}\email{anders@phy.ohiou.edu}
\author{D. R. Phillips}\email{phillips@phy.ohiou.edu}
\affiliation{Department of Physics and Astronomy, 
Ohio University, Athens, OH 45701}

\begin{abstract}
The reaction $\pi^-d\to nn\gamma$ is calculated in chiral perturbation
theory so as to facilitate an extraction of the neutron-neutron
scattering length ($a_{nn}$).  We include all diagrams up to $O(Q^3)$.
This includes loop effects in the elementary $\pi^-p\to\gamma n$
amplitude and two-body diagrams, both of which were ignored in
previous calculations.  We find that the chiral expansion for the
ratio of the quasi-free (QF) to final-state-interaction (FSI) peaks in
the final-state neutron spectrum converges well.  Our third-order
calculation of the full spectrum is already accurate to better than
5\%.  Extracting $a_{nn}$ from the shape of the entire $\pi^-d\to
nn\gamma$ spectrum using our calculation in its present stage would
thus be possible at the $\pm0.8$~fm level. A fit to the FSI peak only
would allow an extraction of $a_{nn}$ with a theoretical uncertainty
of $\pm 0.2$ fm. The effects that contribute to these error bars are
investigated.  The uncertainty in the $nn$ rescattering wave function
dominates.  This suggests that the quoted theoretical error of
$\pm0.3$~fm for the most recent $\pi^-d\to nn\gamma$ measurement may
be optimistic.  The possibility of constraining the $nn$ rescattering
wave function used in our calculation more tightly---and thus reducing
the error---is briefly discussed.
\end{abstract}


\maketitle

\section{Introduction}
In QCD, charge symmetry (CS) is a symmetry of the Lagrangian under the
exchange of the up and down quarks~\cite{MNS}.  This symmetry has many
consequences at the hadronic level, where it translates into, \EG, the
invariance of the strong nuclear force under the exchange of protons
and neutrons.  However, CS is broken by the different masses of the up
and down quarks and thus the strong interaction manifests charge
symmetry breaking (CSB).  The different electromagnetic properties of
the up and down quarks also contribute to CSB.  An important
consequence of the first CSB effect (strong CSB) is that the neutron is 
heavier than the proton, since if CSB was only an electromagnetic effect 
the proton would be heavier and prone to decay.  
This would make our world very different, since big-bang nucleosynthesis 
is dependent on the relative proton and neutron abundances.

While there are a number of pieces of experimental evidences for 
CSB~\cite{MNS}---including recent results in $dd\to\alpha\pi^0$ at 
IUCF~\cite{IUCFCSB} and $np\to d\pi^0$ at TRIUMF~\cite{Allena}---one of the 
most fundamental to nuclear physics is the difference between the 
neutron-neutron ($a_{nn}$) and proton-proton ($a_{pp}$) scattering lengths.
The scattering lengths parameterize the nucleon-nucleon interaction at low 
relative energies through the effective range expansion.
In this low-energy region the ($^1S_0$) phase shift $\delta_0$ can be 
expressed as
\begin{equation}
  p \cot\delta_0 = -\frac{1}{a}+\frac12 r_0p^2,
\label{eq:ER}
\end{equation}
where $p$ is the center-of-momentum (c.m.) relative nucleon momentum and 
$r_0$ the effective range.
This expansion is reliable for $p\alt150$~MeV/$c$.

Since there are no free neutron targets it is very difficult to make a
direct measurement of the $^1S_0$ neutron-neutron scattering length,
though proposals to do so have been made.  The latest of these
suggests using the pulsed nuclear reactor YAGUAR in Snezhinsk,
Russia~\cite{yaguar}.  However, the more common---and so far more
successful---approach is to rely on suitable reactions involving two free
neutrons and corresponding theoretical calculations to extract $a_{nn}$ from 
indirect data.  By choosing the
kinematics carefully one can detect the neutrons in a low-energy
relative $S$-wave which can be accurately described by the effective
range expansion (\ref{eq:ER}).  The currently accepted value
\begin{equation}
a_{nn}=-18.5\pm 0.3\ {\rm fm}
\label{eq:annvalue}
\end{equation} 
is deduced from the break-up reaction $nd\to nnp$ and the pion radiative 
capture process $\pi^-d\to nn\gamma$.  
The first reaction is, however, complicated by the possible presence of
three-body forces, but even after they are taken into account there
are significant disagreements between values extracted by the two
techniques.  A recent $nd$ break-up experiment reports
$a_{nn}=-16.1\pm0.4$~fm~\cite{nd1}, \IE, more than five standard
deviations from the standard value~(\ref{eq:annvalue}).  
The result of Ref.~\cite{nd1} is also in disagreement with another $nd$ 
experiment that claims $-18.7\pm0.6$~fm~\cite{nd2}.
Earlier data had an even larger spread, see Ref.~\cite{Slaus} for a review.  
Since the proton-proton scattering length is
$a_{pp}=-17.3\pm0.4$~fm (after corrections of electromagnetic
effects~\footnote{There is a small electromagnetic correction
($-0.3$~fm) to $a_{nn}$~\protect\cite{MNS}, which for the rest of this
paper will be ignored.}), there is even uncertainty about the sign
of the difference $a_{pp}-a_{nn}$.  The more negative $a_{nn}$ is
favored by nuclear structure calculations, where the small (but
important) CSB piece of the AV18 potential is fitted to reproduce
$a_{pp} - a_{nn}$ with $a_{nn}=-18.5$~fm~\cite{Wi95}.  The
binding-energy difference between $^3{\rm H}$ and $^3{\rm He}$, which
has a small contribution from CSB effects, is then very accurately
reproduced.  This would not occur were $a_{pp} - a_{nn}$ to take the
opposite sign~\cite{AV18}.

Because of these issues the accepted value is weighted toward the
$a_{nn}=-18.50\pm0.05(\rm stat.)\pm0.44(syst.)\pm0.30(theory)$~fm
reported by the most recent $\pi^-d\to nn\gamma$ experiment~\cite{LAMPF}.  
The extraction in this is case done by fitting the shape of the neutron 
time-of-flight spectrum using the model of
Gibbs, Gibson, and Stephenson (GGS)~\cite{GGS}.  This model was
developed in the mid-70s and explored many of the relevant mechanisms
and the dependence on various choices of wave functions.  
Gibbs, Gibson, and Stephenson calculated the single-nucleon radiative 
pion capture tree-level amplitude to order $p/M$, and consequently ignored 
the pion loops that would enter at the next chiral order.  
Two-body diagrams were not fully implemented in this model.  
The theoretical error was dominated
by uncertainties in the scattering wave function.
Similar results were obtained in earlier $\pi^-d\to nn\gamma$ experiments 
carried out at the Paul Scherrer Institut (PSI) (then Swiss Institute of 
Nuclear Research)~\cite{Gabioudetal}.
In the PSI experiments, only the FSI peak was fitted, while at LAMPF the entire
spectrum was fitted.
The theoretical work for the PSI results compared the GGS model with work done
by de~T\'eramond and collaborators~\cite{deTeramond}.
The latter used a dispersion relation approach for the final state interaction,
with a theoretical error of the order 0.3~fm, \IE, similar to GGS.

In this paper we recalculate the $\pi^-d\to nn\gamma$ reaction using
chiral perturbation theory ($\chi$PT).  The one-body and two-body
mechanisms are thus consistent and the constraints of chiral
symmetry are respected, which is of crucial importance in this
threshold regime.  At third order, $O(Q^3)$, all the amplitudes of the
previous calculation are included, as well as pion loops and three
pion-rescattering diagrams.  Additional advantages of the chiral power
counting are that it gives a clearly defined procedure to estimate the
theoretical error and provides a systematic and consistent way to
improve the calculation if needed.
In this first paper we establish the machinery necessary for a precise 
extraction of the $nn$ scattering length.
We isolate the sources of the largest remaining errors and suggest means for 
their reduction, which should make it possible to reach the desired high 
precision in future work.

The paper is organized as follows.  In Sec.~\ref{sec:layout} we will
develop the main ingredients of our calculation: the Lagrangian, the
explicit forms of the one- and two-body amplitudes, and a
description of our wave functions.  The numerical results are
presented in Sec.~\ref{sec:results} together with our estimate of the
theoretical error.  We conclude in Sec.~\ref{sec:end}.

\section{Layout of calculation}
\label{sec:layout}
The LAMPF experiment~\cite{LAMPF} used stopped pions, captured into
atomic orbitals around the deuteron.  The subsequent radiative decay
occurs for pionic $s$-wave orbitals only.  Thus the c.m.\ and
laboratory frames coincide and the pion momentum is vanishingly small.
The neutron time-of-flight distribution of the c.m.\ $\pi^-d$ decay
width (with the photon and one neutron detected) can be expressed as
\begin{equation}
  \frac{d^2\Gamma}{dt_1d\theta_3} = \frac{1}{2(2\pi)^3M_{\pi d}t_1}
  \frac{p_1^3E_1 k\sin\theta_3}{M_{\pi d}-E_1-p_1\cos\theta_3}
  \frac{1}{3}\sum_{\rm pols.}|\mathcal{M}|^2,
\end{equation}
where $t_1$, $p_1$, and $E_1$ are the time-of-flight, momentum, and
energy of the detected neutron, $\theta_3$ is the supplement of the
angle between this neutron and the photon, and $k$ is the photon
momentum.  Here the sum is over deuteron and photon polarizations
and $M_{\pi d}$ is the mass of the deuteron-pion bound
system, which to a very good approximation is given by the sum of the
pion ($\mu$) and deuteron ($M_d$) masses.

\begin{figure}[t]
  \includegraphics{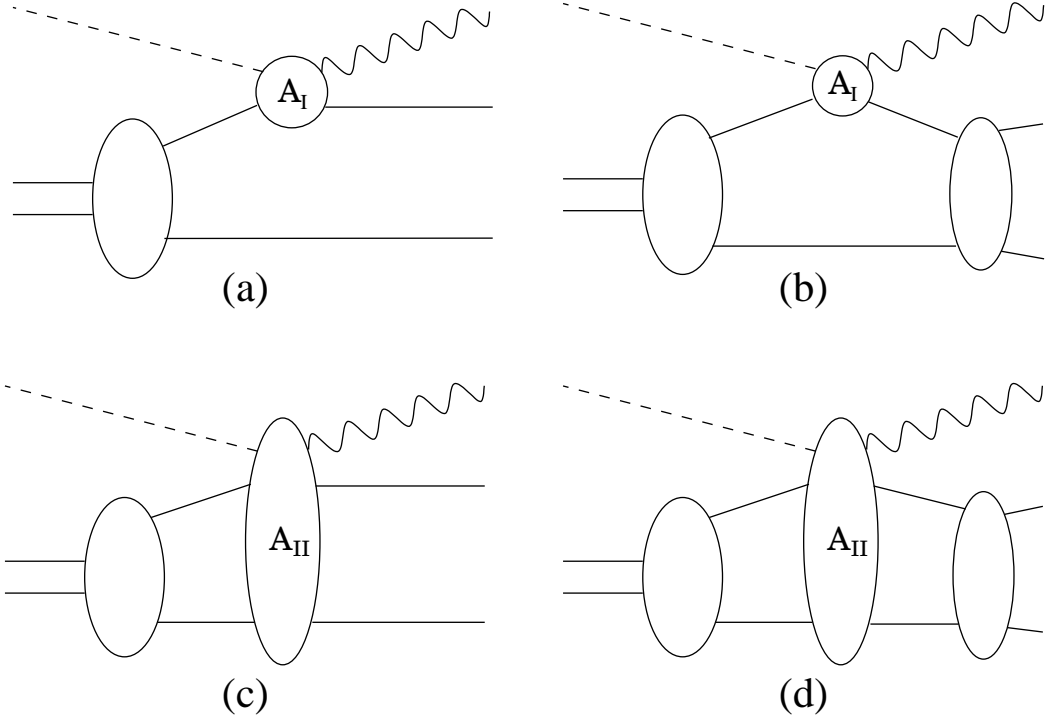}
\caption{Generic diagrams of the contributions to $\pi^-d\to nn\gamma$.
Diagram (a) and (b) are the one-body contributions, without and with FSI.
Diagram (c) and (d) are the corresponding for the two-body currents.
The amplitudes $\mathcal{A}_{\rm I}$ and $\mathcal{A}_{\rm II}$ are described 
in the text.}
\label{fig:gendiag}
\end{figure}

The matrix element $\mathcal{M}$ is the sum of four interfering parts,
the quasi-free (QF) one-body, the one-body with final state
interaction (FSI), and the two-body contributions, with and without
FSI.  These can be symbolized by the generic diagrams shown in
Fig.~\ref{fig:gendiag}.  In this first calculation we restrict FSI to
$S$-waves only, and subtract the plane-wave from the scattering wave
function and include it in the QF contribution.

\begin{figure}[t]
\includegraphics{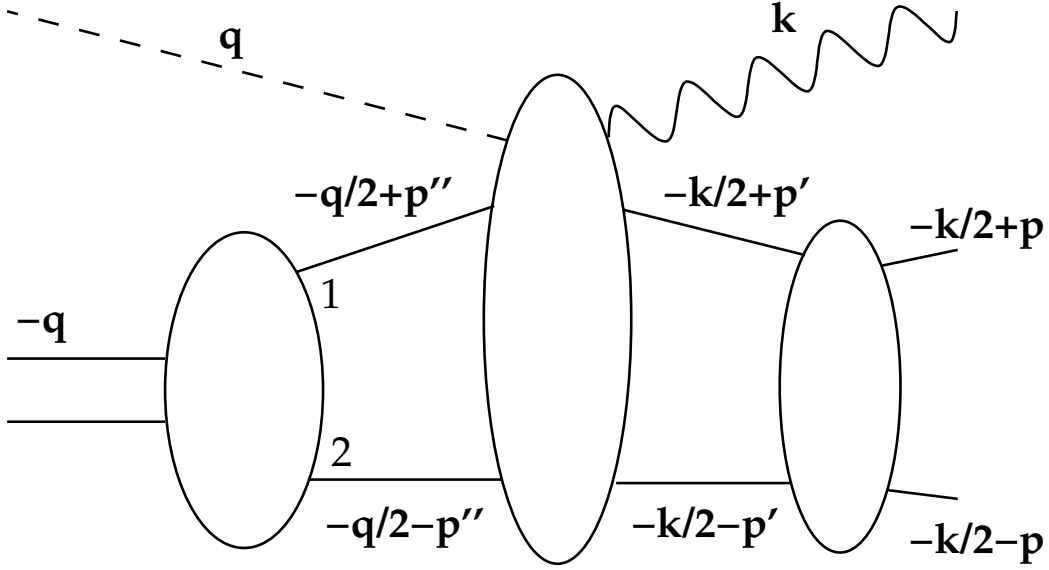}
\caption{Basic diagram for $\pi^-d\to nn\gamma$, defining kinematic variables.}
\label{fig:kin}
\end{figure}

We will derive the matrix elements for $\gamma nn\to\pi^-d$ rather
than $\pi^-d\to nn\gamma$ in order to reduce the possibility of
relative phase errors when using the $\gamma n\to\pi^-p$ amplitudes.
The $\pi^-d\to nn\gamma$ decay rate can of course then be obtained by
detailed balance.  An explicit expression for the matrix
element in terms of $\pi d$ atomic ($\Phi_{\pi d}$) and deuteron
($\varphi_d$) wave functions and the pion-photon amplitude
$\mathcal{A}$ is given by
\begin{eqnarray}
  \mathcal{M}(\gamma nn\to\pi^-d) & = & \int 
  \frac{d^3qd^3p'd^3p''}{(2\pi)^9}\frac{M\sqrt{2E_{\pi d}}}
       {\sqrt{E_1E_22E_\pi}}
  \Phi^\ast_{\pi d}(0;\qi{})\varphi^\ast_{d}(-\qi{};\bp'')
  \mathcal{A}
  \Psi_{\bf -k}(\bp',\bp),
\end{eqnarray}
where $\bp$, $\bp'$, and $\bp''$ are the initial, intermediate, and final 
relative momenta of the two nucleons (for $\gamma nn\to\pi^- d$), while 
$\qi{}$ is the pion c.m.\ momentum and $E_x$ the energy of the indicated 
particle.  
The meaning of these kinematic variables can also be inferred from
Fig.~\ref{fig:kin}.
Here $\Psi_{\bf -k}(\bp',\bp)$ is the $nn$ scattering wave function at total 
momentum ${\bf -k}$, normalized such that
\begin{equation}
  \int \frac{d^3q}{(2\pi)^3}
  \Psi^{\ast}_{\bf -k}(\bp',\qi{})\Psi_{\bf -k}(\bp,\qi{}) = 
  (2\pi)^3\delta^3(\bp'-\bp).
\end{equation}
Its Fourier transform is given by
\begin{equation}
  \Psi_{\bf -k}(\bp',\bp) = \int\frac{d^3r}{(2\pi)^3}\expup{-i\bp'\cdot\ri{}}
  \Psi_{\bf -k}(\ri{},\bp),
\end{equation}
where 
$\Psi_{\bf -k}(\bp,\ri{})=\sum_l(2l+1)i^l\frac{v_l(r)}{r}P_l(\phat\cdot\rhat)$,
which could be compared to the plane wave expansion 
$\expup{i\bp\cdot\ri{}}=\sum_l(2l+1)i^lj_l(pr)P_l(\phat\cdot\rhat)$.
The form of $v_l(r)$ will be discussed in Sec.~\ref{sec:wf}.

The pion-photon amplitude $\mathcal{A}$ can be separated into the one- and 
two-body amplitudes $\mathcal{A}_{\rm I}$ and $\mathcal{A}_{\rm II}$;
\begin{eqnarray}
    \mathcal{A} & = & (2\pi)^3\frac{E}{M}
    \delta^{(3)}(\tilde\bp\pm\frac{\ki{}-\qi{}}{2}) 
    \mathcal{A}_{\rm I}(\ki{},\qi{})+
    \mathcal{A}_{\rm II}(\pm\tilde\bp,\ki{},\qi{}),
\end{eqnarray}
where the upper(lower) sign is for interaction on nucleon 1(2) of the deuteron 
and $\tilde\bp=\bp'-\bp''$.
The one- and two-body matrix elements are then 
\begin{eqnarray}
  \mathcal{M}_{\rm I}(\gamma nn\to\pi^-d) & = & \int \frac{d^3qd^3p'}{(2\pi)^6}
  \frac{E\sqrt{2E_{\pi d}}}{\sqrt{E_1E_22E_\pi}}
  \Phi^\ast_{\pi d}(0;\qi{})
  \varphi^\ast_{d}(-\qi{};\bp'\pm\frac{\bf k-q}{2})
  \mathcal{A}_{\rm I}(\ki{},\qi{})
  \Psi_{-\bf k}(\bp',\bp), \nonumber \\
   \mathcal{M}_{\rm II}(\gamma nn\to\pi^-d) & = & 
   \int \frac{d^3qd^3p'd^3p''}{(2\pi)^9}
   \frac{M\sqrt{2E_{\pi d}}}{\sqrt{E_1E_22E_\pi}}
   \Phi^\ast_{\pi d}(0;\qi{})
   \varphi^\ast_{d}(-\qi{};\bp'')
   \mathcal{A}_{\rm II}(\pm\tilde\bp,\ki{},\qi{}) \nonumber \\
   & \times & \Psi_{-\bf k}(\bp',\bp). 
\end{eqnarray}

In configuration space the matrix elements are given by
\begin{eqnarray}
  \mathcal{M}_{\rm I}^{\rm FSI} & = & \sum\Phi_{\pi d}(0) 
  \sqrt{\frac{M_{\pi d}}{\mu}}
  \int dr\frac{d\Omega_r}{\sqrt{4\pi}}
  [S_0u(r)+S_2(\rhat)w(r)]\expup{\pm\frac{i}{2}(\ki{}-\qi{})\cdot{\bf r}}
  \mathcal{A}_{\rm I}(\ki{},\qi{}) \tilde{v}_0(r)\chi_0, \\
  \mathcal{M}_{\rm I}^{\rm QF} & = & \sum\Phi_{\pi d}(0) 
  \sqrt{\frac{M_{\pi d}}{\mu}}
  \int rdr\frac{d\Omega_r}{\sqrt{4\pi}} 
  [S_0u(r)+S_2(\rhat)w(r)]\expup{\pm \frac{i}{2}(\ki{}-\qi{})\cdot{\bf r}}
  \mathcal{A}_{\rm I}({\bf k},\qi{})\expup{i\bp\cdot\ri{}}\chi_S, \\
  \mathcal{M}_{\rm II}^{\rm FSI} & = & \sum\Phi_{\pi d}(0) 
  \sqrt{\frac{M_{\pi d}}{\mu}}
  \int dr \frac{d\Omega_r}{\sqrt{4\pi}}
  [S_0u(r)+S_2(\rhat)w(r)]\mathcal{A}^\pm_{\rm II}({\bf r},{\bf k},\qi{})
  \tilde{v}_0(r)\chi_{0}, \\
  \mathcal{M}_{\rm II}^{\rm PW} & = & \sum\Phi_{\pi d}(0) 
  \sqrt{\frac{M_{\pi d}}{\mu}}
  \int rdr \frac{d\Omega_r}{\sqrt{4\pi}}
  [S_0u(r)+S_2(\rhat)w(r)]\mathcal{A}^\pm_{\rm II}({\bf r},{\bf k},\qi{})
  \expup{\pm i\bp_i\cdot\ri{}}\chi_S, 
\end{eqnarray}
where ${\bf p}_{1,2}=-\frac{1}{2}{\bf k}\pm{\bf p}$, 
$\mathcal{A}^\pm_{\rm II}({\bf r},{\bf k},{\bf q})=\int 
\frac{d^3\tilde{p}}{(2\pi)^3}\expup{-i\tilde\bp\cdot{\bf r}}
\mathcal{A}_{\rm II}(\pm\tilde\bp,{\bf k},{\bf q})$, 
and $\tilde{v}_0(r)$ is the subtracted scattering wave function as defined in 
Sec.~\ref{sec:wf}.
The sums are over the two nucleons.
In these expressions $\Phi_{\pi d}(0)=(\mu_{\pi d}\alpha)^{3/2}/\sqrt{\pi}$ is 
the pion-deuteron atomic $s$-orbital wave function evaluated at the origin 
with $\mu_{\pi d}$ the reduced $\pi d$ mass, while 
$S_0=-\frac{1}{\sqrt{2}}\bsigma\cdot\beps_d^\dagger$ and
$S_2(\rhat)=\frac{1}{2}(3\bsigma\cdot\rhat\,\rhat\cdot\beps_d^\dagger-
\bsigma\cdot\beps_d^\dagger)$ 
are the $S$- and $D$-wave spin structures of the deuteron, $\beps_d^\dagger$ 
being the deuteron polarization vector.
The $\chi_S$'s are the neutron-neutron spin wave functions for spin $S$.

In the following subsections we will derive explicit expressions for the 
amplitudes and show how the coordinate-space wave functions are obtained.

\subsection{Power counting}
\label{sec:EFT}
We start from the relativistic Lagrangian
\begin{eqnarray}
  \mathcal{L} & = & {\cal L}_{\pi N}^{(1)} + {\cal L}_{\pi N}^{(2)} +
{\cal L}_{\pi \pi}^{(2)}; \nonumber \\
{\cal L}_{\pi N}^{(1)} & = & \bar N\left[i\gamma\cdot D_N-M
-\frac{1}{4f_\pi^2}\epsilon^{abc}\tau^c\pi^a\gamma\cdot\partial\pi^b
-\frac{g_A}{2f_\pi}(\gamma\cdot D^{ab} \pi^a)\tau^b\gamma_5\right] N, 
\nonumber\\ 
{\cal L}_{\pi N}^{(2)} & = & -\bar
N\left[\frac{e(\kappa_0+\kappa_1\tau^3)}{8M}\sigma^{\mu\nu}F_{\mu\nu}\right]N,
\nonumber \\ 
{\cal L}_{\pi \pi}^{(2)}&=& \frac12D_\mu^{ab}\pi^a D^{\mu
cb}\pi^c-\frac12\mu^2\bpi^2,
\label{eq:Lagrange}
\end{eqnarray}
where $D^\mu_N = \partial^\mu-ieQ_NA^\mu$ and 
$D^{\mu ab}_\pi=\delta^{ab}\partial^\mu-ieQ^{ab}_\pi A^\mu$ are the covariant 
derivatives for the nucleon and pion ($e<0$), $Q_N=\frac12(1+\tau^3)$ and 
$Q^{ab}_\pi=i\epsilon^{ab3}$ are the electric charge isospin operators 
(with $a$/$b$ isospin indices of incoming/outgoing pion), and 
$\kappa_{0,1}=\kappa_p\pm\kappa_n$ ($\kappa_p=1.793$ and $\kappa_n=-1.913$)
represent the nucleon anomalous magnetic moments. 
The electromagnetic field tensor is given by 
$F_{\mu\nu}=\partial_\nu A_\mu-\partial_\mu A_\nu$ and
$\sigma^{\mu\nu}=\frac{i}{2}[\gamma^\mu,\gamma^\nu]$ as usual.

The Lagrangian (\ref{eq:Lagrange}) is organized according to the
number of powers of small momenta $Q$ (here $e$ counts as one ``small''
momentum). The chiral order of any graph to any amplitude involving
nucleons, together with pions and photons with energies of order
$\mu$ can be assessed by multiplying the $Q$-scaling factors of the
individual units of the graph by one another. These factors are as
follows:
\begin{itemize}
\item Each vertex from ${\cal L}^{(n)}$ contributes $Q^n$.

\item Each nucleon propagator scales like $1/Q$ (provided that
the energy flowing through the nucleon line is $\sim\mu$);

\item Each pion propagator scales like $1/Q^2$;

\item Each pion loop contributes $Q^4$;

\item Graphs in which two nucleons participate in the reaction
acquire an extra factor of $Q^3$.
\end{itemize}
In practice tree-level relativistic graphs must be calculated,
and then expanded in powers of $p/M$ in order to establish
contact with the usual heavy-baryon formulation of chiral
perturbation theory. The expressions for loop graphs that we use
are also computed in heavy-baryon $\chi$PT, and so we do not
need to employ any special subtraction schemes to remove
pieces of loop integrals which scale with positive powers 
of the nucleon mass~\cite{BL99,Fu03}.

Note that we will employ the Coulomb gauge in all calculations.

\subsection{One-body amplitudes}
\label{sec:oneB}
\begin{figure}[t]
\includegraphics{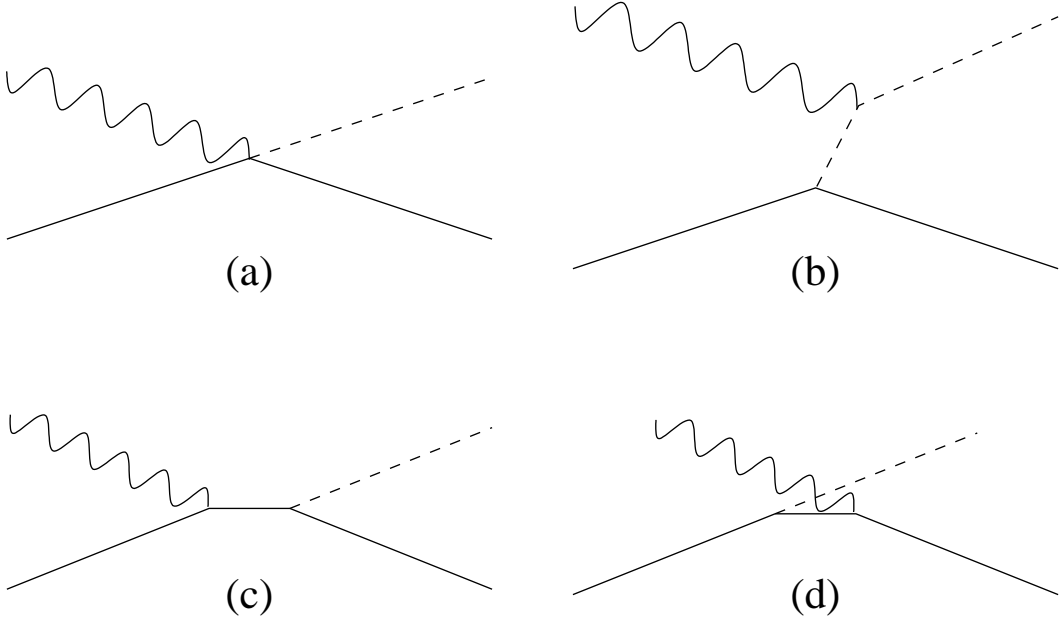}
\caption{Single-nucleon pion-photon diagrams relevant for pion photoproduction.
In this and all other figures a solid line represent a nucleon, 
a dashed line a pion and a wavy line a photon.}
\label{fig:one}
\end{figure}

\begin{figure}[t]
\includegraphics{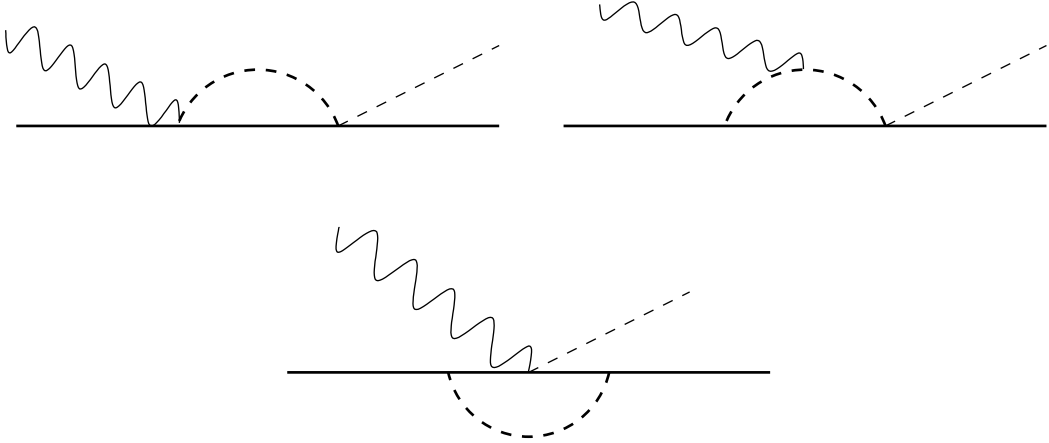}
\caption{Pion loops at NNLO for $\gamma N\to \pi N$ in Coulomb gauge. 
Note that not all time orderings are shown.}
\label{fig:piloops}
\end{figure}

There are four basic one-body diagrams, shown in Fig.~\ref{fig:one}:
the Kroll-Ruderman (KR) term (a), the pion pole (b), and the $s$- and 
$u$-channel nucleon pole terms (c) and (d).
They can be calculated directly from a non-relativistic reduction of
the relativistic Lagrangian or from the amplitudes of heavy-baryon
$\chi$PT (HB$\chi$PT)~\cite{ulfreview}.  In addition, there are also
pion-loop corrections at $O(Q^3)$ as shown in Fig.~\ref{fig:piloops}.
The loop corrections, together with the corresponding counterterms from 
${\cal L}_{\pi N}^{(3)}$~\cite{EM} have already been calculated in the
$\gamma N$ c.m.~frame to $O(Q^3)$ for radiative pion capture on a
nucleon~\cite{Fearing} using Coulomb gauge.  The low-energy constants
(LECs) from the third-order chiral Lagrangian were fitted to
experiment, yielding an excellent description of the near-threshold
data.

The third-order piece of the one-body amplitudes of
Ref.~\cite{Fearing} can hence be taken as is, without introducing any
new unknown parameters, and combined with our evaluation to $O(Q^2)$
of the tree-level diagrams in Fig.~\ref{fig:one} using the
relativistic Lagrangian (\ref{eq:Lagrange})~\footnote{The amplitudes
of Ref.~\protect\cite{Fearing} are based on the third-order
heavy-baryon Lagrangian of Ecker and
Moj\v{z}i\v{s}~\protect\cite{EM}. There are some differences between
the results obtained in this way and those found when the tree-level
relativistic amplitude for $\gamma n \rightarrow \pi^- p$ is expanded
to relative order $p^2/M^2$. Thus the tree-level terms we find at $O(Q^3)$
are slightly different to those listed in Ref.~\cite{Fearing}, but
this can be accounted for by a redefinition of the LECs in ${\cal
L}_{\pi N}^{(3)}$.  This redefinition only affects the $O(Q^3)$
(NNLO) photoproduction amplitude and thus---at the order we consider
here---it is not relevant to the boost or other issues associated with
embedding the $\gamma n \rightarrow \pi^- p$ amplitude in the $A=2$
system.}. In order to incorporate these amplitudes in the two-body
system they should be evaluated at the relevant subthreshold
kinematics, corrected for a boost to the overall rest frame, and
corrected for off-shell effects. These issues will all be discussed
below.

The full one-body amplitude is given (in Coulomb gauge) by
\begin{eqnarray}
  \mathcal{A}_{\rm I}(\gamma N\to\pi N) & = & 
  F_1(E_\pi,x)i\bsigma\cdot\beps_\gamma+
  F_2(E_\pi,x)\bsigma\cdot\qhat\,\bsigma\cdot(\khat\times\beps_\gamma)+
  F_3(E_\pi,x)i\bsigma\cdot\khat\,\qhat\cdot\beps_\gamma \nonumber \\
  & + & F_4(E_\pi,x)i\bsigma\cdot\qhat\,\qhat\cdot\beps_\gamma,
\label{eq:Fis}
\end{eqnarray}
where the $F_i$ are the Chew-Goldberger-Low-Nambu (CGLN) amplitudes~\cite{CGLN}
and $\beps_\gamma$ is the photon polarization vector.
The isospin channels are separated as
\begin{equation}
  F_i^a(E_\pi,x) = F_i^{(-)}(E_\pi,x)i\epsilon^{a3b}\tau^b+
  F_i^{(0)}(E_\pi,x)\tau^a+F_i^{(+)}(E_\pi,x)\delta^{a3},
\label{eq:isospin}
\end{equation}
where $a$ is the pion isospin index.  For $\gamma n\to\pi^-p$ this
implies that $F_i=\sqrt{2}[F_i^{(0)}-F_i^{(-)}]$.  The $F_i$'s of
Ref.~\cite{Fearing} are evaluated with the pion energy $E_\pi$ and
photon-pion cosine $x=\khat\cdot\qhat$ in the $\pi^-p\to\gamma n$ rest
frame.  In our case $\qi{}=0$, $E_\pi=\mu$, and $x$ is undetermined.
Thus only $F_1$ survives, the other spin amplitudes being proportional
to the pion momentum.  In charged pion photo-production $F_1$ is
dominated by the KR contribution [Fig.~\ref{fig:one}(a)].

\subsubsection{Subthreshold extrapolation}
\label{sec:subthreshold}
If the $\pi^-p\to\gamma n$ process was completely free, the CGLN
amplitudes should be evaluated at the pion threshold $E_\pi=\mu$.
However, since the proton is bound in the deuteron, the $\pi^-p$
energy is actually less than $\mu$, which means that we must extrapolate
to the sub-threshold regime. To do this we need
a prescription to calculate the invariant two-body energy
$s_{\pi^-p}$.  The pion and photon energy in the $\pi^-p\to\gamma n$
rest frame can then be calculated using the well-known relations
\begin{eqnarray}
  E_\pi^\ast & = & \frac{s_{\pi^-p}-m_p^2+\mu^2}{2\sqrt{s_{\pi^-p}}}, \\
  \omega^\ast & = & \frac{s_{\pi^-p}-m_n^2}{2\sqrt{s_{\pi^-p}}}.
\end{eqnarray}
The energy available to the $\pi^- p$ subsystem, $s_{\pi^- p}$ would
seem to be different depending on whether FSI or QF kinematics are
considered.  Furthermore, there are two QF situations, \IE, the
detected neutron can originate from the one-body vertex or it can be
the spectator. In fact, the first case is overwhelmingly favored by
the kinematics of the LAMPF experiment and is also the one closest to
threshold.  The second, spectator, scenario is suppressed by kinematics, so
even though it is further from threshold and so results in a larger
shift in $s_{\pi^- p}$, any correction resulting from this shift is
small compared to other, included, effects.

For the QF kinematics where the detected neutron originates from the
one-body vertex the rest frame coincides, by definition, with the
overall $\gamma nn$ c.m. But, in the FSI region one has to make a
choice.  The invariant energy of the $\pi^-p\to\gamma n$ system can be
established from
\begin{equation}
  s_{\pi^- p} = (M_d+\mu)^2+m_n^2-2(M_d+\mu)\epsilon_s,
\label{eq:spipspect}
\end{equation}
where $\epsilon_s=\sqrt{m_n^2+p_s^2}$ is the energy of
the spectator nucleon.  We choose to assume that the spectator nucleon
is on-shell and that its typical momentum $p_s$ can be estimated
through calculating the expectation value $\langle p_s^2\rangle$
between initial and final state wave functions. Using the $S$-state of
the deuteron only, the average is given by
\begin{eqnarray}
  \langle p_s^2\rangle & = & \frac{k^2}{4}-
  \frac{\int dr(MB-p^2+2MV_{SS})u(r)j_0(\frac{kr}{2})\tilde{v}_0(r)}
       {\int dr u(r)j_0(\frac{kr}{2})v(r)}.
\label{eq:pssq}
\end{eqnarray}
We then use free kinematics for the one-body amplitudes in the QF
region, and the formulas (\ref{eq:spipspect}) and (\ref{eq:pssq})
to calculate the energy at which the one-body amplitude should
be evaluated in the FSI peak. 
The one-body amplitudes are then calculated using $E_\pi^\ast$ according 
to the different kinematics of the QF and FSI configurations. 
The theoretical uncertainty due to this procedure is assessed in 
Sec.~\ref{subsec:extrap}.

\subsubsection{Boost corrections}
In general the $\gamma n\to\pi^-p$ rest frame does not coincide with
the overall c.m., so we have to adjust the $F_i$ for boost effects.
The boost corrections can be calculated by replacing the $\gamma n\to
\pi^-p$ rest frame kinematics by the overall $\gamma nn$ c.m.\
kinematics in the evaluation of the one-body amplitudes.  This changes
the incoming and outgoing nucleon momenta, but not the photon and pion
momenta.  The Coulomb gauge condition $\epsilon_\gamma^0=0$ is
retained.  From the Lagrangian~(\ref{eq:Lagrange}) one can then
deduce the following boost corrections for the reduced amplitudes, up
to order $Q^2/M^2$ for the $\gamma N\to\pi N$ reactions;
\begin{eqnarray}
  \Delta F_1^{(0)}(E_\pi) & = & \frac{eg_A}{2f_\pi} 
  \frac{-(E_\pi\bp_n\cdot\khat+E_\pi^2)}{2M^2}(\mu_p+\mu_n),
\label{eq:corrF10} \\
  \Delta F_1^{(-)}(E_\pi) &=& \frac{eg_A}{2f_\pi}
  \frac{E_\pi\bp_n\cdot\khat+E_\pi^2}{M^2},
  \label{eq:corrF1m}
\end{eqnarray}
where $\bp_n$ is the outgoing nucleon momentum ($=-\ki{}$ in the $\gamma n$ 
rest frame, which makes these amplitudes vanish).
As before we have assumed Coulomb gauge, $\qi{}=0$, and the same isospin 
designations as in Eq.~(\ref{eq:isospin}).
These corrections should thus be added to the amplitudes of Eqs.~(\ref{eq:Fis})
and (\ref{eq:isospin}) as given in~\cite{Fearing}, except for an overall 
factor of $M/4\pi\sqrt{s}$, which is included in the phase 
space in our formalism.
There are also terms with new spin-momentum structures:
\begin{eqnarray}
  G^{(0)}(E_\pi) & = & \frac{eg_A}{2f_\pi} 
  \frac{iE_\pi\bp_n\cdot\beps_\gamma\bsigma\cdot\khat}{2M^2}
  (\mu_p+\mu_n-1), \\
  G^{(-)}(E_\pi) & = & \frac{eg_A}{2f_\pi}\left(
  \frac{E_\pi\bp_n\cdot(\khat\times\beps_\gamma)}{2M^2}(\mu_p-\mu_n+\frac12)-
  \frac{i\bp_n\cdot\beps_\gamma\bsigma\cdot(2\bp_n+E_\pi\khat)}{M^2}
  \right),
\end{eqnarray}
which will also vanish in the limit $\bp_n\to-\ki{}$.
In the case of non-vanishing pion momentum, additional terms will show up.
One would expect that the first term of $G^{(-)}$ should give the largest 
contribution since $\mu_p-\mu_n+\frac12=5.2$ is a big number.
However, because of the particular kinematics of the present problem, 
$\bp_n\approx-\ki{}$ and the triple scalar product 
$\bp_n\cdot(\ki{}\times\beps_\gamma)\approx E_\pi^2\sin\theta_3$ 
with $\theta_3=0.075$.
Thus, ultimately this piece of $G^{(-)}$ is very small because of the 
kinematics.
Similarly $\bp_n\cdot\beps_\gamma\approx E_\pi\sin\theta_3$.
(Additionally, only one of the photon polarizations can contribute.)
In fact it turns out that the new spin-momentum structures have a negligible 
effect on the pion-photon amplitude and the only possible relevant boost 
corrections come from the terms in Eqs.~(\ref{eq:corrF10}) and 
(\ref{eq:corrF1m}).
In the actual calculations the subthreshold value $E_\pi^\ast$ 
and not $E_\pi=\mu$ was used in evaluating these.

\subsection{Two-body amplitudes}
\label{sec:twoB}
\begin{figure}[t]
\includegraphics{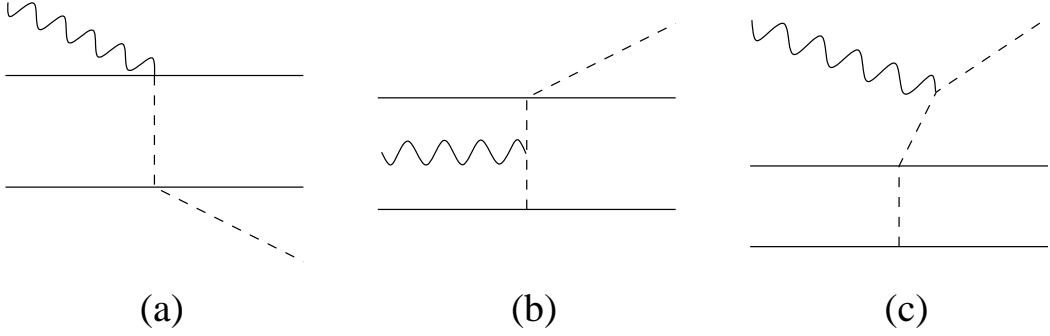}
\caption{The third order pion rescattering diagrams relevant for 
$\pi^-d\to nn\gamma$.
They are shown in order of expected importance as explained in the text.}
\label{fig:two}
\end{figure}

At third order there are three pion-exchange diagrams, displayed in 
Fig.~\ref{fig:two}.
Of these, the first one (a) is expected to give the largest contribution since 
its propagator is coulombic, \IE, behaves like $1/\qi{}^2$, where $\qi{}$ is 
the momentum of the exchange pion~\cite{BLvK}.
This is because (for our kinematics) the pion energy is transferred completely 
to the photon and vanishes against the pion mass in the propagator, thus 
putting the pion effectively on-shell.
It has been argued in the literature that when the intermediate nucleon-nucleon
state for this diagram (as interpreted in time-ordered perturbation theory) is 
Pauli-allowed, corrections due to nucleon recoil need
to be taken into account~\cite{Baru}.
However, in our case the intermediate nucleon pair is in a triplet-isotriplet
state implying a relative $P$-wave, which is Pauli-suppressed.
The recoil correction evaluated in Ref.~\cite{Baru} is thus small.
The second graph (b) has an extra pion propagator which is off-shell, reducing 
the magnitude of this diagram.
The third two-body amplitude (c) has two off-shell pion propagators and is 
hence suppressed compared to the other two.
More importantly, this diagram is further suppressed since in Coulomb gauge it 
is proportional to the (vanishingly small) pion momentum.

The two-body amplitudes, corresponding to the diagrams of Fig.~\ref{fig:two} 
(a-c), are
\begin{eqnarray}
  \mathcal{A}_{{\rm II}a}(\tilde\bp,{\bf k},{\bf q}) & = & \frac{eg_A}{2f_\pi}
  \frac{(-2iE_\pi)}{4f_\pi^2}
  \left[\frac{\tau_1^a\tau_2^3\bsigma_1\cdot\beps_\gamma}
    {(\tilde\bp+\frac{\ki{}+\qi{}}{2})^2}
    +\frac{\tau_1^3\tau_2^a\bsigma_2\cdot\beps_\gamma}
    {(\tilde\bp-\frac{\ki{}+\qi{}}{2})^2} \right], \\
  \mathcal{A}_{{\rm II}b}(\tilde\bp,{\bf k},{\bf q}) & = & \frac{eg_A}{2f_\pi}
  \frac{4iE_\pi}{4f_\pi^2}
  \left[\frac{\tau_1^a\tau_2^3\bsigma_1\cdot\left(\tilde\bp-
      \frac{\ki{}-\qi{}}{2}\right)
      \beps_\gamma\cdot(\tilde\bp+\qi{})}
    {(\tilde\bp+\frac{\ki{}+\qi{}}{2})^2[\mu^2+(\tilde\bp-
	\frac{\ki{}-\qi{}}{2})^2]}
    \right.\nonumber \\ & + & \left.
    \frac{\tau_1^3\tau_2^a\bsigma_2\cdot\left(\tilde\bp+
      \frac{\ki{}-\qi{}}{2}\right)
      \beps_\gamma\cdot(\tilde\bp-\qi{})}
    {(\tilde\bp-\frac{\ki{}+\qi{}}{2})^2[\mu^2+(\tilde\bp+
	\frac{\ki{}-\qi{}}{2})^2]}
    \right], \\
  \mathcal{A}_{{\rm II}c}(\tilde\bp,{\bf k},{\bf q}) & = & \frac{eg_A}{2f_\pi}
  \frac{2E_\pi-\omega}{4f_\pi^2}
  \frac{(\tau_1^a\tau_2^3-\tau_1^3\tau_2^a)i\beps_\gamma\cdot\qi{}}
       {\omega(E_\pi-qy)}
  \left[ \frac{\bsigma_1\cdot(\tilde\bp-\frac{\ki{}-\qi{}}{2})}
    {\mu^2+(\tilde\bp-\frac{\ki{}-\qi{}}{2})^2}
    +\frac{\bsigma_2\cdot(\tilde\bp+\frac{\ki{}-\qi{}}{2})}
    {\mu^2+(\tilde\bp+\frac{\ki{}-\qi{}}{2})^2}\right], \nonumber \\
\end{eqnarray}
where $y=\khat\cdot\qhat$ is the pion-photon cosine in the overall c.m.
In configuration space (for $\qi{}=0$) the two-body amplitudes can be expressed
as
\begin{eqnarray}
  \mathcal{A}_{{\rm II}a}({\bf r},{\bf k},{\bf q=0}) & = & 
  \frac{eg_A}{8f_\pi^3}\frac{-2iE_\pi}{4\pi r}
  \left(\tau_1^a\tau_2^3\bsigma_1\cdot\beps_\gamma
  \expup{\frac{i}{2}{\bf k\cdot r}}
  +\tau_1^3\tau_2^a\bsigma_2\cdot\beps_\gamma
  \expup{-\frac{i}{2}{\bf k\cdot r}}  \right), \\
  \mathcal{A}_{{\rm II}b}({\bf r},{\bf k},{\bf q=0}) & = & 
  -\frac{eg_A}{8f_\pi^3}\frac{2E_\pi}{4\pi}
  \tau_1^a\tau_2^3 \int d\alpha \expup{-\tilde\mu r} 
  \left[\beps_\gamma\cdot\rhat 
  \left( \bsigma_1\cdot[(1-\alpha){\bf k}+i(\tilde\mu+\frac{1}{r})\rhat]
  \expup{i(\frac12-\alpha){\bf k\cdot r}} \right.\right.\nonumber \\
  & + & \left.
  \bsigma_2\cdot[(1-\alpha){\bf k}-i(\tilde\mu+\frac{1}{r})\rhat]
  \expup{-i(\frac12-\alpha){\bf k\cdot r}}\right) \nonumber \\
  & - & \left. \frac{i}{r}
  (\bsigma_1\cdot\beps_\gamma\expup{i(\frac12-\alpha){\bf k\cdot r}}-
  \bsigma_2\cdot\beps_\gamma\expup{-i(\frac12-\alpha){\bf k\cdot r}}) 
  \right], \\
  \mathcal{A}_{{\rm II}c}({\bf r},{\bf k},{\bf q=0}) & = & 0,
\end{eqnarray}
where $\tilde\mu^2=\alpha(\mu^2+\omega^2)-\alpha^2\omega^2$.
These expressions agree with the ones derived in Ref.~\cite{BLvK} after the 
sign correction of Ref.~\cite{Be97}.

\subsection{Matrix elements}
\label{sec:M}
The full matrix elements for the QF amplitudes, projected on spin-0 and spin-1 
final states are, after taking the trace over nucleon spins and isospins
\begin{eqnarray}
  \mathcal{M}_0 & = & Ci \left[\beps_d^\dagger\cdot\beps_\gamma(F_1-xF_2)+
    \khat\cdot\beps_d^\dagger\qhat\cdot\beps_\gamma(F_2+F_3)+
    \qhat\cdot\beps_d^\dagger\qhat\cdot\beps_\gamma F_4\right] f(p_2) 
  \nonumber \\
  & + & \frac{3Ci}{\sqrt2} \phat_2\cdot\beps_d^\dagger
  \left[\phat_2\cdot\beps_\gamma(F_1-xF_2)+\phat_2\cdot\khat\qhat\cdot
    \beps_\gamma(F_2+F_3)+
    \phat_2\cdot\qhat\qhat\cdot\beps_\gamma F_4\right] g(p_2)+(2\rightarrow1), 
\nonumber \\
  \mathcal{M}_1 & = & C\left\{\beps_d^\dagger\cdot
  (\beps_\gamma\times\beps_{nn})F_1-
  \left[\qhat\cdot\beps_d^\dagger\beps_{nn}-\qhat\cdot\beps_{nn}\beps_d^\dagger
    +\beps_d^\dagger\cdot\beps_{nn}\qhat\right]\cdot(\khat\times\beps_\gamma)
  F_2 \right.
  \nonumber \\ & + & \left.
  \beps_d^\dagger\cdot(\khat\times\beps_{nn})\qhat\cdot\beps_\gamma F_3+
  \beps_d^\dagger\cdot(\qhat\times\beps_{nn})\qhat\cdot\beps_\gamma F_4\right\}
  f(p_2) \nonumber \\
  & + & \frac{3C}{\sqrt2}\phat_2\cdot\beps_d^\dagger\left[ 
  \phat_2\cdot(\beps_\gamma\times\beps_{nn})F_1
  -\left(\phat_2\cdot\qhat\beps_{nn}-\qhat\cdot\beps_{nn}\phat_2
  +\phat_2\cdot\beps_{nn}\qhat\right)\cdot(\khat\times\beps_\gamma)F_2 \right. 
  \nonumber \\
  & + & \left. \phat_2\cdot(\khat\times\beps_{nn})\qhat\cdot\beps_\gamma F_3
  +\phat_2\cdot(\qhat\times\beps_{nn})\qhat\cdot\beps_\gamma F_4 \right] 
  g(p_2)-(2\rightarrow1) ,
\end{eqnarray}
where $\beps_{nn}$ is the polarization vector of a spin-1 neutron pair, 
\begin{eqnarray}
  C & = & \sqrt{4\pi}\Phi_{\pi d}(0)\sqrt{\frac{M_{\pi d}}{\mu}},\\
  f(p) & = & \int \, r \, dr\left[u(r)j_0(pr)-{{1}\over{\sqrt 2}}w(r)j_2(pr)
    \right],\\ 
  g(p) & = & \int \,r \, dr w(r) j_2(pr).
\end{eqnarray}
The corresponding spin-0 FSI matrix element is easily obtained by the
replacement $\bp_2\to\ki{}$ and letting 
\begin{eqnarray}
f(k)&=&\int \, dr \, \left[u(r)j_0\left(\frac{kr}{2}\right)-\frac{1}{\sqrt 2}
w(r)j_2\left(\frac{kr}{2}\right)\right]\tilde{v}_0(r),\\
g(k)&=&\int \, dr \, w(r)j_2\left(\frac{kr}{2}\right)\tilde{v}_0(r).
\end{eqnarray} 
The symmetrization ($2\rightarrow1$) is then equivalent to an overall
factor of two in the spin-0 FSI matrix element.  Similar expressions
can be derived for the two-body amplitudes and for higher partial
waves.

\subsection{Wave functions}
\label{sec:wf}
It is possible to calculate quite accurate deuteron and
nucleon-nucleon scattering wave functions from the well-established
asymptotic states. By using data extracted from the Nijmegen
phase-shift analysis~\cite{NijmPWA} as well as a one-pion-exchange
potential we ensure that the behavior of the wave function at 
$r\agt\frac{1}{m_\pi}$ is correct.  This yields wave functions that are
consistent with those obtained from $\chi$PT potentials at leading
order~\cite{weinNN}. In order to be fully consistent with the
$O(Q^3)$, or NNLO, operators we have derived here one should of course
include $O(Q^2)$ corrections to the $NN$ potential, \IE, incorporate
at least the ``leading'' chiral two-pion exchange
(TPE)~\cite{Or96,Ka97,Ep99,Re99}.  This will be done in future work.

\subsubsection{The deuteron wave function}
The deuteron wave function at large distances is described by the
asymptotic $S$- and $D$-state wave functions:
\begin{eqnarray}
  u^{(0)}(r) & = & A_S\expup{-\gamma r}, \\
  w^{(0)}(r) & = & \eta A_S\left(1+\frac{3}{\gamma r}+
  \frac{3}{(\gamma r)^2}\right) \expup{-\gamma r},
\end{eqnarray}
where $\gamma=\sqrt{MB}=45.70223(9)$~MeV/$c$ [$B=2.224575(9)$~MeV], 
$A_S=0.8845(8)$~fm$^{-1/2}$ is the asymptotic normalization, and 
$\eta=0.0253(2)$ the asymptotic $D/S$ ratio~\cite{Nijmd}.
The (un-regulated) deuteron wave functions $u(r)$ and $w(r)$ can be obtained 
 from the asymptotic ones and the radial Schr\"odinger equation by integrating 
in from $r=\infty$~\cite{PC}
\begin{eqnarray}
  u(r) & = & u^{(0)}(r)-M\int_r^\infty dr' G_0(r',r)[V_{SS}(r')u(r')+
    V_{SD}(r')w(r')], \nonumber \\
  w(r) & = & w^{(0)}(r)-M\int_r^\infty dr' G_2(r',r)[V_{DS}(r')u(r')+
    V_{DD}(r')w(r')].
\label{eq:inteqd}
\end{eqnarray}
Here, $G_{0/2}(r',r)$ is the $S$- and $D$-wave Green function (propagator) and 
$V_{L'L}$ the standard projections of the Yukawa OPE potential:
\begin{eqnarray}
  V_{SS} & = & -f^2\frac{\expup{-\mu r}}{r}, \nonumber \\
  V_{SD}=V_{DS} & = & -2\sqrt2 f^2
  \frac{\expup{-\mu r}}{r} \left(1+\frac{3}{\mu r}+\frac{3}{(\mu r)^2}\right), 
  \nonumber \\
  V_{DD} & = & -f^2\frac{\expup{-\mu r}}{r}
  +2 f^2\frac{\expup{-\mu r}}{r}
  \left(1+\frac{3}{\mu r}+\frac{3}{(\mu r)^2}\right),
\end{eqnarray}
where $f^2=0.0750(5)$ is the $\pi NN$ coupling constant 
squared~\cite{NijmpiNN}.
The coupled integral equation (\ref{eq:inteqd}) is solved using standard
numerical techniques.

The integrated wave functions are divergent at small distances, reflecting that
short-range physics has been ignored.
Instead of trying to model this piece as done in many phenomenological $NN$ 
potentials, we choose to regulate it by matching with a spherical well solution
at $r=R_d$.
This procedure is motivated by the fundamental EFT hypothesis that results 
should not be sensitive to the behavior at small $r$.
If it is, there are some short-distance physics that need to be included in
the calculation, \IE, that the parametrization is incomplete.
This hypothesis can be tested by varying the cut-off $R_d$ over some sensible 
range.
A thorough discussion of the boundary between long- and short-distance 
physics along these lines can be found in the lectures by 
Lepage~\cite{Lepage:1997cs}.

The $D$-wave ($w$) is matched at the boundary $R_d$ by the continuity of the 
logarithmic derivative, which determines the depth of the well.
The $S$-wave is matched assuming continuity and that the deuteron wave function
is normalized to unity.
The matching condition is then
\begin{equation}
  1-\int_{R_d}^\infty dr u^2(r)-\int_0^\infty dr w^2(r)=\int_0^{R_d} dr u^2(r),
\end{equation}
where the left-hand side is calculated numerically and the right-hand
side analytically.

\begin{figure}[t]
\includegraphics[width=135mm]{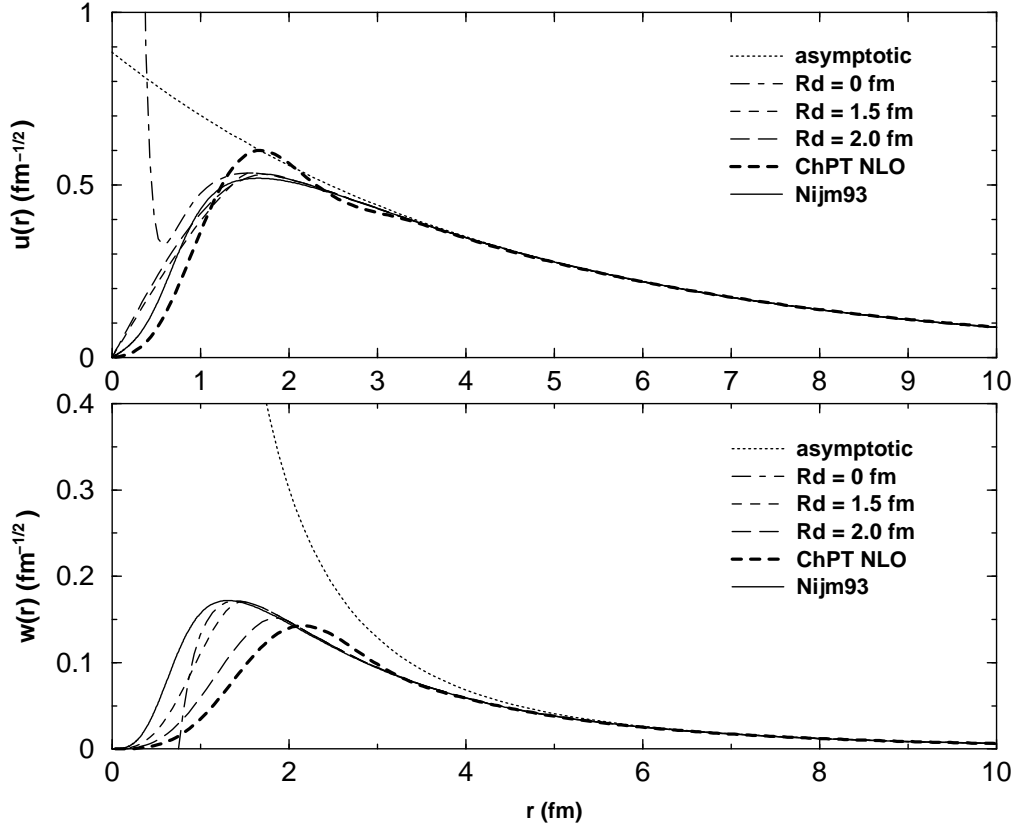}
\caption{The $S$- (top panel) and $D$-state (bottom panel) parts of the 
deuteron wave functions.
Our wave functions are labeled by the choice of matching radius $R_d$.
These are compared to the NLO chiral wave function of
Ref.~\protect\cite{Ep99} and Nijm93~\protect\cite{nijmPot}.}
\label{fig:wfsd}
\end{figure}

In Fig.~\ref{fig:wfsd} these wave functions are compared to each other, to
the modern chiral NLO wave function of Epelbaum~\EA~\cite{Ep99}, and to the 
wave function of the Nijm93 potential~\cite{nijmPot}.
Choosing $R_d$ to be in the range 1.5--2~fm gives wave functions that are very 
close to the high-precision wave functions on the market.
Note that the chiral wave function (thick dashed line in Fig.~\ref{fig:wfsd})
deviates considerably from the potential model wave function (solid line)
between 1.5 and 2.0~fm.
We take this as an indication that short-range $NN$ dynamics are at play even
at distances as large as 2.0~fm.
Thus we vary our matching point $R_d$ (and $R_{nn}$ below) between 1.5 and 
2~fm, where the lower limit is set because $R_d$ cannot be reduced much further
without the interaction becoming non-Hermitian.
It is possible to use the asymptotic wave functions (dotted lines in
Fig.~\ref{fig:wfsd}) in the calculation if the matrix element has the
necessary factors of $r$ to cancel the divergences of the wave functions 
as $r\to0$.  
Such an approach is similar in spirit to the pionless effective field theory 
[EFT($\not\!\pi$)] and gives analytic expressions for the matrix element.

\subsubsection{The $nn$ scattering wave function}
The scattering wave function can be calculated in a similar way. 
However, the asymptotic state is now described by the phase shift according to
\begin{equation}
  \Psi_{\bf -k}({\bf r},\bp) \sim \frac{v^{(0)}_0(r)}{r} = 
  \frac{\expup{i\delta_0}\sin(pr+\delta_0)}{pr},
\end{equation}
where $\delta_0$ is calculated from Eq.~(\ref{eq:ER}) with given values of $p$,
$a_{nn}$, and $r_0$.
In the limit of vanishing momentum this wave function reduces to $1-a/r$ as it 
should.
If higher partial waves can be neglected the only integral equation we need is
the one for the $^1S_0$ channel, which is
\begin{equation}
  v_0(r) = v^{(0)}_0(r)-M\int_r^\infty dr' 
  \widetilde{G}_0(r',r)V_{SS}(r')v_0(r'),
\end{equation}
where $\widetilde{G}_0(r',r)$ is the free $S$-wave two-body propagator.
This wave function is regularized at $r=R_{nn}$ by matching the logarithmic 
derivative of a spherical well solution.
The depth of the spherical well is hence energy dependent, a treatment related 
to the energy-dependent potential used by Beane~\EA,~\cite{towards}.
As for the deuteron we vary $R_{nn}$ between 1.5 and 2~fm in order to
test our sensitivity to short-range dynamics.

\begin{figure}[t]
\includegraphics*[width=140mm]{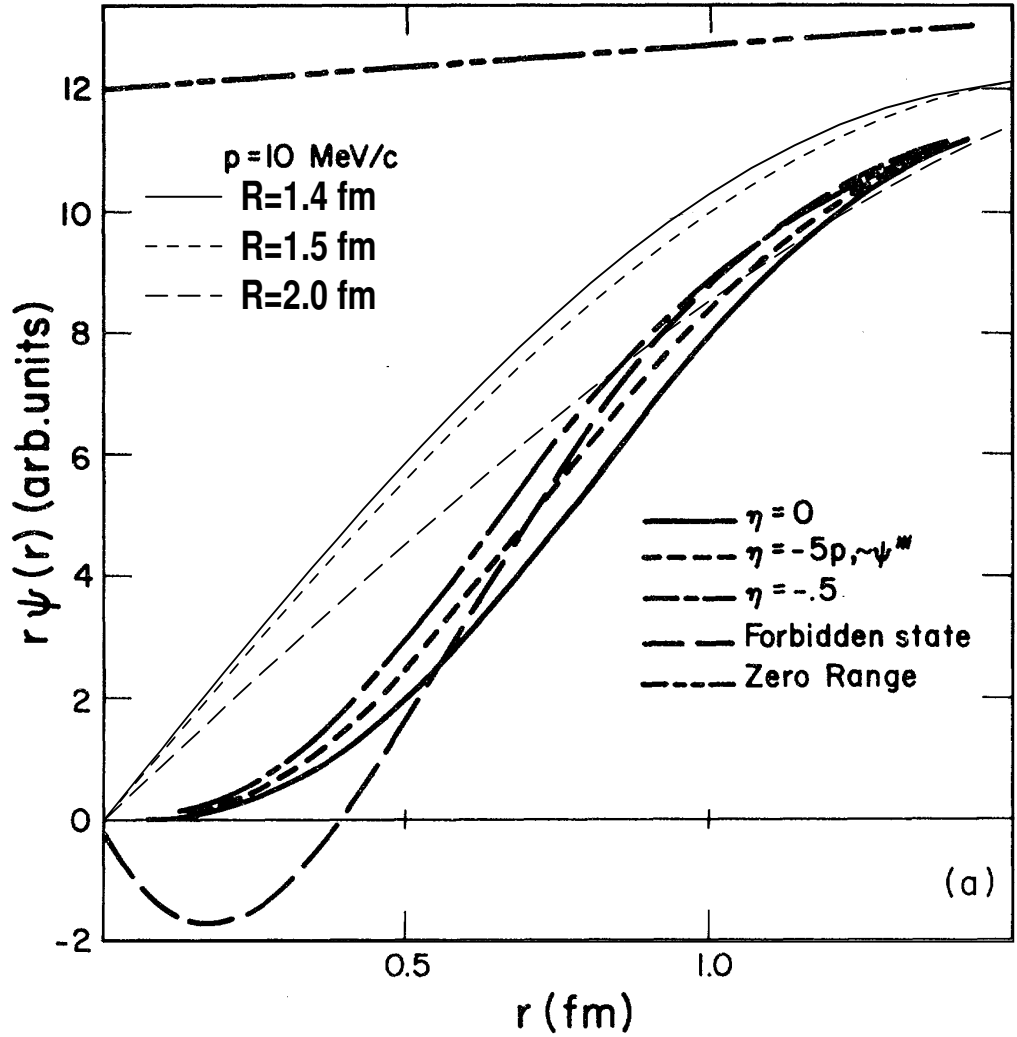}
\caption{Our $^1S_0$ $nn$ scattering wave functions (thin lines) at 
$p=10$~MeV/c and for varying $R_{nn}$ as indicated.
Here $a_{nn}=-16$~fm and $r_0=2.8$~fm.
A comparison is made with the wave functions of the GGS model (thick lines).
The latter are explained in~\protect\cite{GGS}, from which reference the figure
was adapted.
Reprinted figure with permission from 
W.~R.~Gibbs, B.~F.~Gibson, and G.~J.~Stephenson, Jr.,
Phys\ Rev.\ C {\bf 11}, 90 (1975). 
Copyright (1975) by the American Physical Society.}
\label{fig:nnwf}
\end{figure}
In Fig.~\ref{fig:nnwf} our $nn$ wave functions are plotted together
with the wave functions used by GGS.  The wave functions are quite
similar in that they tend to the same asymptotic limit (the curve
labeled Zero Range) for larger $r$.  Thus they are very close to each
other for $r\agt1.5$~fm.  There are, however, a few important
differences between the $nn$ wave functions in the two calculations:
Firstly, the GGS wave functions have been derived using the Reid
soft-core potential (RSC) (with the old larger value for the $\pi NN$
coupling constant) for the long-range part, while we used one-pion
exchange only.  
This explains the slight difference in the size of $v(r)$ at $r=1.4$~fm.
Secondly, GGS match at a fixed value of $R_{nn}=1.4$~fm, while we vary 
$R_{nn}$.
Thirdly, our wave function uses a spherical well solution ($\sin\kappa r$) 
and GGS assume a polynomial of fifth order, where the magnitude and first
two derivatives vanish at $r=0$ and are matched to the RSC solution at 
$r=R_{nn}$.  
The assumption of vanishing derivatives is equivalent
to using a hard-core potential at short distances, while many chiral 
potentials have a softer behavior for small $r$ (see, \EG,~Ref.~\cite{Ep99}).  
[The different shapes of the GGS wave functions were obtained by adding an
extra term $\eta r^3(r-R_{nn})^3/pr$ to the short-range piece of their wave
function.]
The combined effect of all this is that our wave function
has most variation around $r=1$~fm, while the GGS wave functions
varies most around $\sim0.7$~fm.  As we will see later, these
differences have a strong influence on the assessment of the size
of the theoretical error in the extracted $a_{nn}$.

An obvious improvement of our calculation would be to use wave
functions whose short-range behavior is constrained by other
observables, thus reducing the uncertainty.  We could also compare to
wave functions of modern high-precision potentials,
\EG,~Refs.~\cite{Wi95,nijmPot}, or the recent N$^3$LO chiral
potentials~\cite{EM03,Ep04}.  Another extension would be to include
higher partial waves.

In the actual calculations we subtract off the plane-wave $S$-wave contribution
$j_0(pr)$ from the scattering wave function [$\tilde{v}_0=v_0-rj_0(pr)$] and 
then calculate the full plane-wave (QF or PW) contribution using 
$\expup{i\bp\cdot\ri{}}$ without partial-wave decomposition.

\section{Results}
\label{sec:results}

\subsection{Convergence}
\begin{figure}[t]
\includegraphics{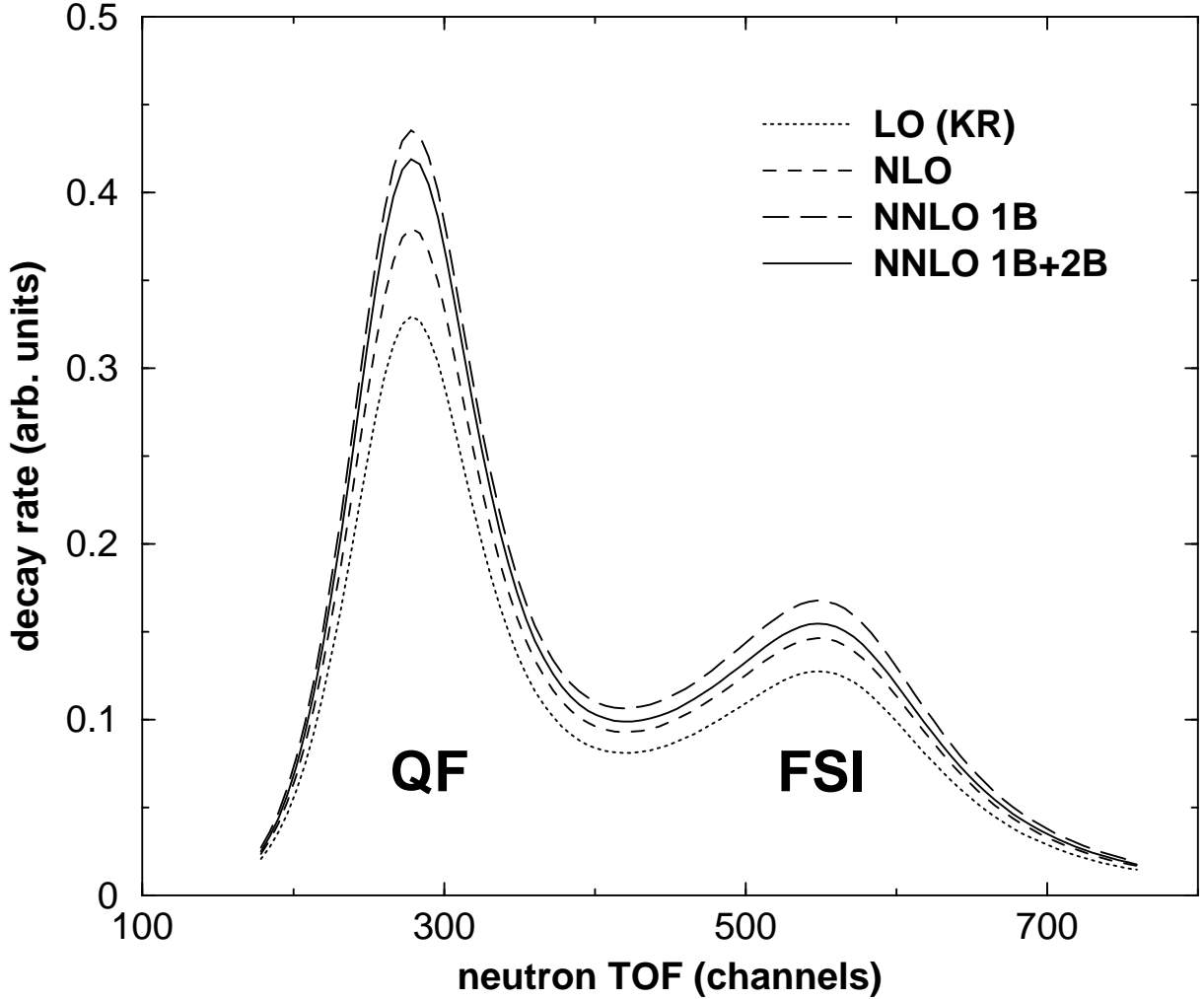}
\caption{Time-of-flight distribution for the $\pi^-d\to nn\gamma$ decay rate.
This and all following spectra are calculated assuming $a_{nn}=-18$~fm, 
$r_0=2.75$~fm, and $\theta_3=0.075$~rad, unless otherwise indicated.
The plot shows the contributions from the LO KR (dotted line), NLO 
one-body (short-dashed line), NNLO one-body (long-dashed line), and NNLO one- 
and two-body (solid line) amplitudes.
The two peaks are labeled QF and FSI from their dominant contributions.}
\label{fig:KR}
\end{figure}

The calculated differential decay width is shown in Fig.~\ref{fig:KR}
for the LO KR term only and with the NLO and NNLO one- and two-body
amplitudes added in succession.
The spectrum shows two separate peaks, labeled QF and FSI from the
dominant contributions that give rise to them.  
It is clear that the LO curve is very similar to the full 
calculation and that the corrections of higher orders mainly affect the 
magnitude, but do have some impact on the shape.  
The evolution is most easily assessed by forming the QF to FSI peak 
ratio at the various orders.  
At LO the ratio is 2.58, at NLO 2.59, at NNLO one-body 2.60, and at full NNLO
2.71.  Thus the LO, NLO, and NNLO one-body results are very close to
each other, but do not contain the full dynamics that the NNLO
two-body amplitudes provide.  For a good quantitative result it is
important to include the full NNLO amplitude---using only one-body
amplitudes would give a wrong answer at this order.

\subsection{Sensitivity to $a_{nn}$}
\label{sec:errorA}

\begin{figure}[t]
\includegraphics{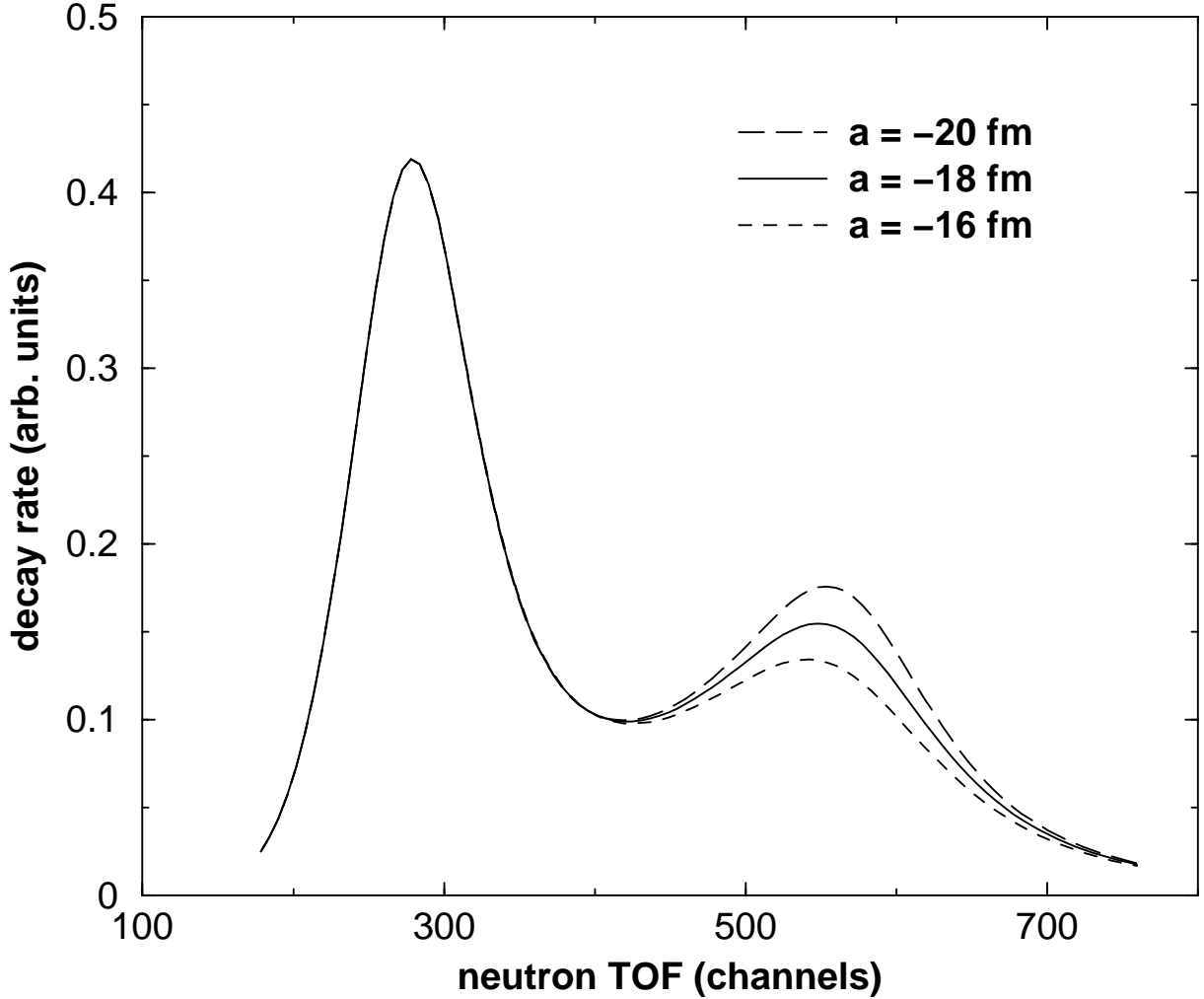}
\caption{Time-of-flight distribution for the $\pi^-d\to nn\gamma$
decay rate.  The spectrum is calculated for different choices of
$a_{nn}$ as indicated.  The curves for $a_{nn}=-20$ and -16~fm have
been (slightly) rescaled to coincide with the QF peak of the
$a_{nn}=-18$~fm curve.}
\label{fig:spectra}
\end{figure}

In Fig.~\ref{fig:spectra} the decay rate is plotted for various
choices of $a_{nn}$.  The curves have been rescaled to coincide at the
QF peak, to facilitate comparison of the relative height of the QF and
FSI peak.  This is done since in the LAMPF experiment~\cite{LAMPF} the
scattering length is extracted by fitting the shape of the spectrum,
not the magnitude of the decay rate.
Note that only the height of the FSI peak changes, the valley between
the two peaks is largely unaffected by the value of $a_{nn}$.

The theoretical error in the extraction of the scattering length has
several sources and they will be investigated and estimated in the
following paragraphs.  The error in the extracted $a_{nn}$ can be
related to the error in the decay rate $\Gamma$ by
\begin{equation}
  \frac{\Delta\Gamma}{\Gamma} = \frac{d\Gamma}{da_{nn}}\frac{a_{nn}}{\Gamma}
  \frac{\Delta a_{nn}}{a_{nn}},
\end{equation}
where the actual calculations (Fig.~\ref{fig:spectra}) give that 
$\frac{d\Gamma}{da_{nn}}\frac{a_{nn}}{\Gamma}=1.21$ at $a_{nn}=-18$~fm.
Thus 
\begin{equation}
\frac{\Delta a_{nn}}{a_{nn}}=0.83\frac{\Delta\Gamma}{\Gamma},
\end{equation}
a result we shall use repeatedly in what follows.

\subsection{Theoretical error bar}
We estimate the theoretical error under the assumption that the entire
time-of-flight spectrum is fitted.
To the best of our knowledge, previous work have only considered fitting the
FSI peak, which limits the kinematics~\cite{GGS,deTeramond,Gabioudetal}.
Because of the large relative momentum in the QF region, this extended analysis
will have significant importance for the size of the error.
We will use a nominal value of $a_{nn}=-18$~fm in our estimate of the error.

\subsubsection{Neglected higher orders in the $\pi^- p \rightarrow
\gamma n$ amplitude}
\label{subsec:extrap}
The present calculation ignore pieces of the $\pi^-p\to\gamma n$ amplitude of 
$O(Q^4)$ or higher, which is thus three orders down from the leading piece of 
$O(Q)$.  
One might think that the error would then be of the order
$(\omega/\Delta)^3$, since the first dynamical effects not explicitly
included in our Lagrangian are associated with $\Delta$-isobar
excitation and so the `high'-energy scale is $\Delta$, the
$\Delta$-nucleon mass difference, rather than $\Lambda_{\chi}$.  This
is supported by the results of Ref.~\cite{Fearing} where the fitted
counter terms had unnaturally large coefficients when expressed in
units of ${\rm GeV}^{-2}$.  
However, in Ref.~\cite{Fearing} the $O(Q^3)$ one-body (single-nucleon 
pion photo-production) amplitude was fitted to actual data for 
$\omega_0=142$~MeV and higher (roughly 10~MeV above threshold).  
The error in our calculation is thus introduced only in our extrapolation 
of the amplitude to a subthreshold energy, denoted by $\omega^\ast$.  
Compared to the leading $O(Q)$ term, this gives a correction
$(\omega_0^3-{\omega^\ast}^3)/\Delta^3\sim4\%$.  
This is a special, very beneficial feature of the pion absorption process: 
since the pion momentum is vanishing there is no angular dependence and the
amplitudes depend only on the photon energy.
This error should include the errors due to uncertainties in the LECs fitted 
in~\cite{Fearing}.
A simple calculation based on the LEC fit errors and the formulas for the CGLN 
amplitudes gives corrections of the order 3\% or smaller, which is in line 
with the above 4\%.

Since the extrapolation photon energy is roughly the same at the QF
and the FSI peak, this error should add roughly equally to both peaks,
which reduces the error on the neutron time-of-flight spectrum to less than the
$\sim4\%$ estimated above, since
now only the shape is fitted.  
The actual calculations confirm this: the spectrum using extrapolated 
amplitudes (see Sec.~\ref{sec:subthreshold}) differs by only 1.1\% 
in the FSI peak from the spectrum with amplitudes evaluated at threshold.  
The corresponding error in $a_{nn}$ is thus 0.95\% or 0.17~fm for 
$a_{nn}=-18$~fm.

\subsubsection{Boost corrections}
The contribution of the boost corrections [Eqs.(\ref{eq:corrF10}) and 
(\ref{eq:corrF1m})] is of the order $\mu^2/2M^2\sim1\%$, but the change occurs 
in the same direction in both peaks.
Thus the relative change is much reduced ($0.14\%$ in 
$\Gamma_{\rm FSI}/\Gamma_{\rm QF}$, \IE, 0.11\% or 0.02~fm in $a_{nn}$) and can
be completely neglected for the present purposes.
After rescaling, the boosted curve cannot be distinguished from the original 
one.
We use the calculated $O(Q^3)$ boost correction of 0.14\% as a conservative 
estimate of the boost error introduced at higher orders.
The boost correction is included in all plots.

\subsubsection{Off-shellness}

\begin{figure}
\includegraphics{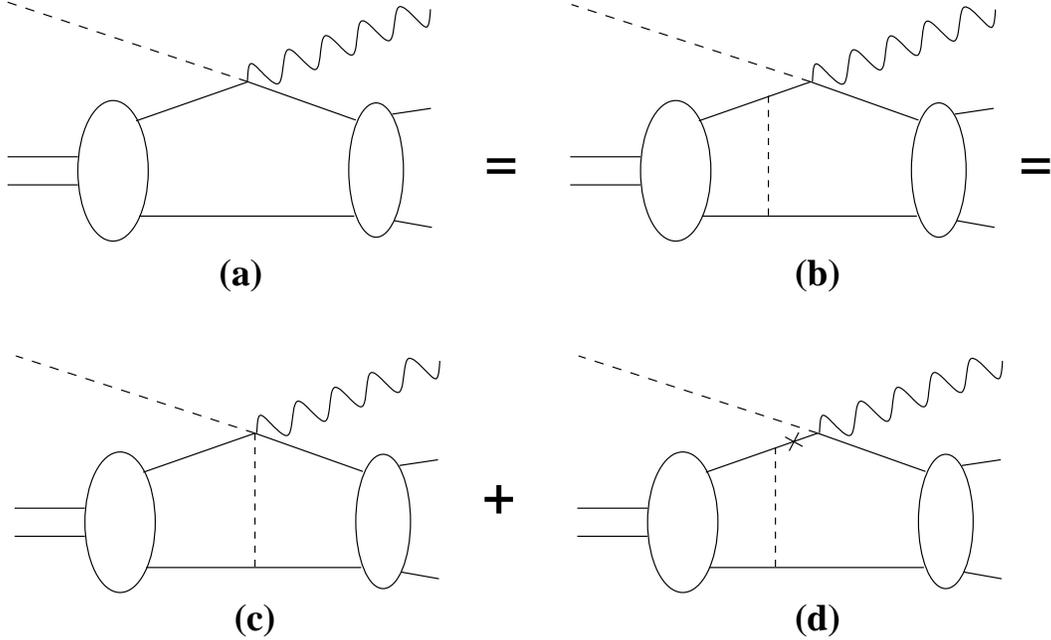}
\caption{The off-shell nucleon in (a) can be taken care of by extracting a 
meson exchange from the deuteron wave function as in (b).
The off-shell part absorbs the closest propagator and becomes the two-body 
diagram (c), splitting off the on-shell amplitude (d).
The cross indicates an on-shell nucleon.}
\label{fig:offshell}
\end{figure}

In calculating the one-body amplitudes we tacitly assumed that both
nucleons were on-shell.  This introduces an ``off-shellness'' error
that should be estimated. It is well known that field transformations
can be employed to trade dependence of the one-body amplitude on the
``off-shellness'' of the nucleon, $p^0 - \frac{{\bf p}^2}{2 M}$, for a
two-body amplitude~\cite{Haag,FH}. This is done as
follows: using field transformations such as those employed in
Refs.~\cite{FS,Fe} the dependence of the $\gamma n \rightarrow \pi^-p$ 
amplitude on the nucleon energy $p^0$ is replaced by dependence on
$\frac{{\bf p}^2}{2 M}$ plus terms in ${\cal L}^{(3)}_{\pi N}$ and
beyond. This means that the one-body amplitude for the photoproduction
process now has no ``off-shell ambiguity'', although we do acquire
additional pieces of the two-body amplitude for the charged-pion
photoproduction process.  (This argument is shown graphically in the
sequence of diagrams in Fig.~\ref{fig:offshell}.)  The new
contribution, depicted in Fig.~\ref{fig:offshell}(c), involves a
$\gamma \pi \pi$ vertex from ${\cal L}_{\pi N}^{(3)}$, and so is
$O(Q^5)$.  This two-body effect is thus $p^2/M^2\sim\mu^2/M^2\sim2\%$
down from the NNLO two-body diagrams, which contribute $7.1\%$ to the
rescaled decay rate (according to Fig.~\ref{fig:KR}). Consequently the
error in $a_{nn}$ from any potential ``off-shell ambiguity'' is
approximately $0.02\times0.071\times0.83=0.12\%$ or 0.02~fm.

\subsubsection{$O(Q^4)$ two-body pieces of the amplitude}
\begin{figure}[t]
\includegraphics{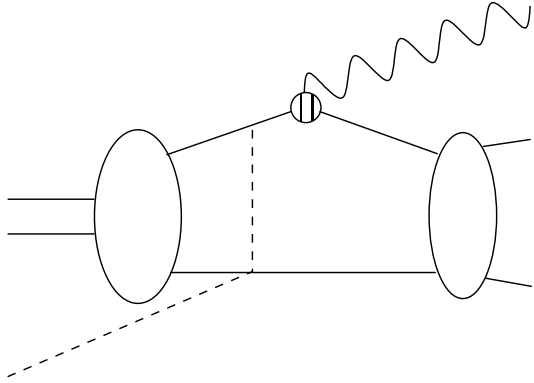}
\caption{Typical two-body operator at $O(Q^4)$. 
The sliced photo-nucleon vertex is from $\mathcal{L}_{\pi N}^{(2)}$.}
\label{fig:oq4tb}
\end{figure}

A larger effect comes from $O(Q^4)$ two-body pieces of the $\gamma nn
\rightarrow \pi^- d$ amplitude, such as the one depicted in 
Fig.~\ref{fig:oq4tb}. 
A naive estimate indicates they should be 
$\sim p/\Lambda_\chi \sim \mu/\Lambda_\chi \sim 20$\% of the 
$O(Q^3)$ two-body diagrams. This estimate is supported
by studies of pion photoproduction to $O(Q^4)$ in $\chi$PT~\cite{Be97}. 
This suggests that $O(Q^4)$ two-body effects are roughly a 
0.7\% effect in $a_{nn}$.

\subsubsection{Error from wave functions}
\label{sec:errorwf}
\begin{figure}[t]
\includegraphics[width=140mm]{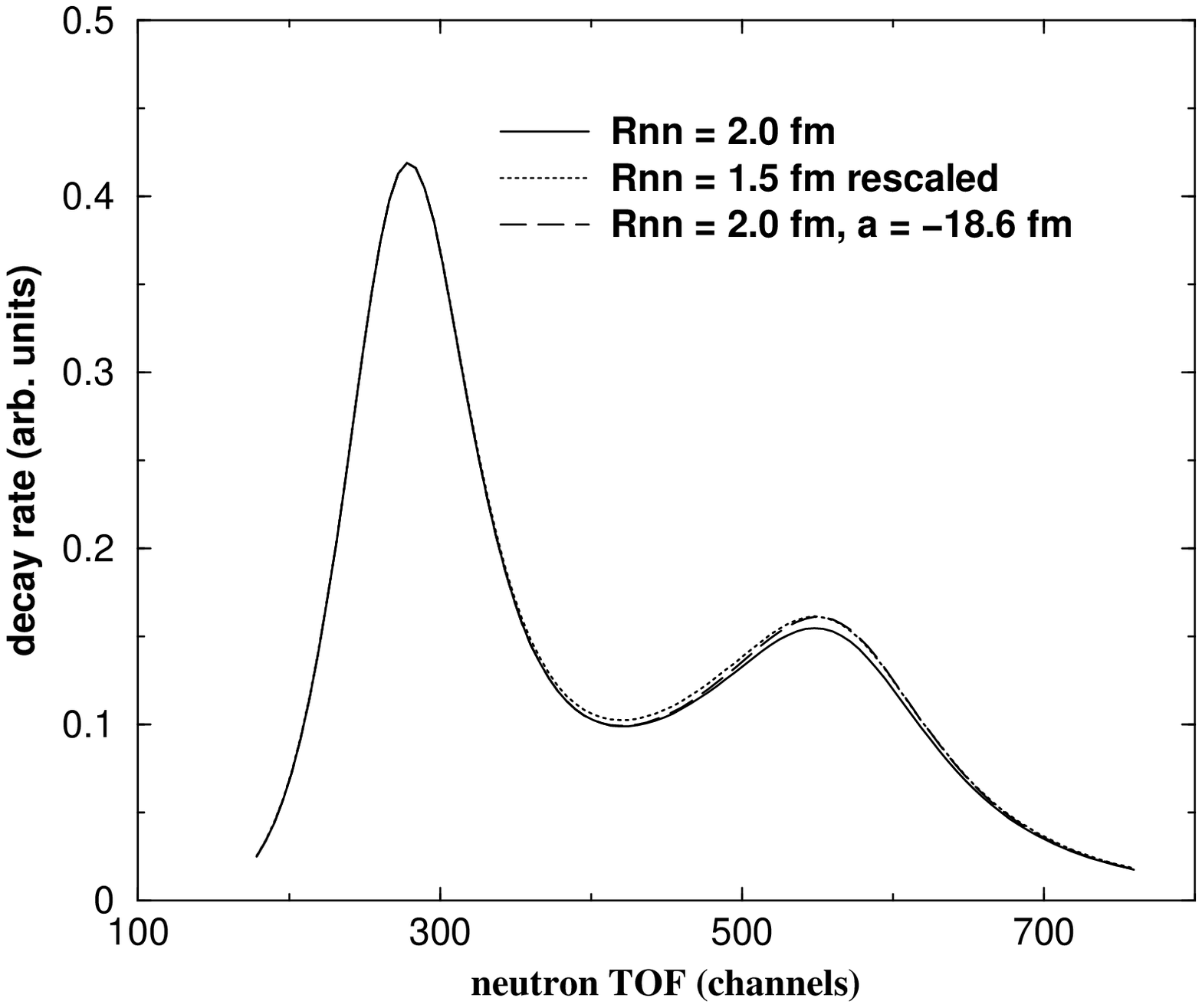}
\caption{Time-of-flight distribution for the $\pi^-d\to nn\gamma$ decay rate.
The spectra are calculated for $a_{nn}=-18$~fm and different 
choices of $R_{nn}$ as indicated.
The case of $R_{nn}=2.0$~fm and $a_{nn}=-18.6$~fm is also plotted.}
\label{fig:spectrawf}
\end{figure}
By changing the matching points $R_d$ and $R_{nn}$ between 1.5 and 2.0
fm, we tested the error introduced by our ignorance of short-distance
physics in the $NN$ wave functions.  This change was significant for
the $nn$ scattering wave functions, as shown in
Fig.~\ref{fig:spectrawf}.  The resulting error in $a_{nn}$ 
turns out to be $-0.6$~fm (3.3\%) or smaller.  A similar spread
was obtained with wave functions calculated from ``high-quality'' $NN$
potentials, \EG~Nijm I and Nijm II~\cite{nijmPot}.  We note that both
these potentials have $\chi^2/{\rm d.o.f.}=1.03$ with respect to the
1993 Nijmegen database and identical $nn$ scattering lengths.  They
differ only in their treatment of the heavy mesons, indicating that
our calculation is sensitive to truly short-range parts of the $NN$
interaction. Our results with the NijmI and NijmII potentials also
suggest that this short-range sensitivity has a greater impact on the
extracted $a_{nn}$ than does our neglect of two-pion exchange. We are
confident that this uncertainty could be considerably reduced by
finding other observables that constrain the wave function, in
particular its short-range behavior.  This will be pursued in future
work.  Note that the change in $R_{nn}$ not only changes the height of
the FSI peak, but also the valley region.  This feature could
potentially be used in a fit procedure to distinguish the $R_{nn}$
dependence from a change in $a_{nn}$.

\begin{figure}[t]
\includegraphics[width=140mm]{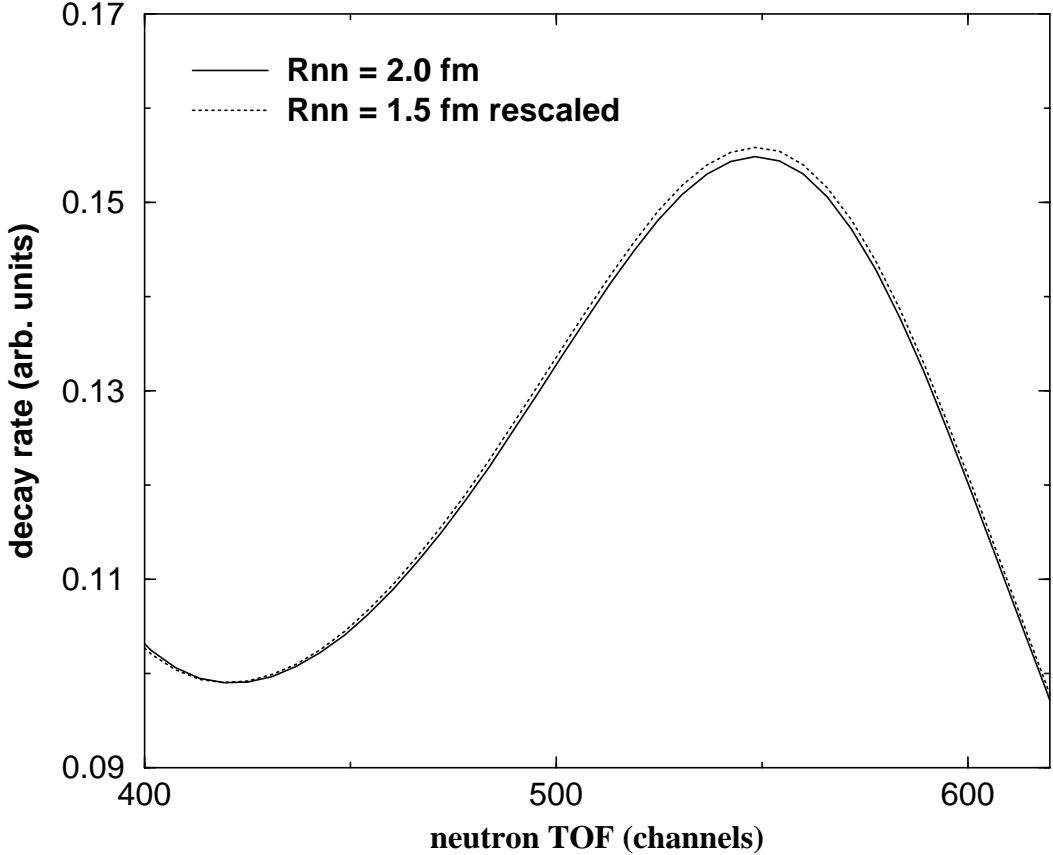}
\caption{Time-of-flight distribution for the $\pi^-d\to nn\gamma$ decay rate
in the region of the FSI peak. 
The spectra are calculated for $a_{nn}=-18$~fm and different 
choices of $R_{nn}$ as indicated.}
\label{fig:FSIspectrawf}
\end{figure}

Indeed, if one focuses only on the FSI peak then the variation in the
spectrum due to the use of different wave functions is significantly
smaller than the one discussed in the previous paragraph. If we adjust
both calculations to agree in the valley region we find that the FSI
peak height only differs by 0.6\%. (See Fig.~\ref{fig:FSIspectrawf}.)
This corresponds to an uncertainty in $a_{nn}$ of $\pm 0.1$ fm.

Meanwhile, effects due to the bound-state wave function chosen are
also small. Changing the deuteron wave function by varying $R_d$ from 
2.0~fm to 1.5~fm would alter the extracted $a_{nn}$ by 0.55\% or
0.10 fm.  Using the Bonn B deuteron wave function instead of the
EFT-motivated wave function yields a $\Delta a_{nn}$ of 0.56\% or
0.10~fm.

\subsubsection{Higher partial waves}
The error from neglecting higher partial waves in the rescattering wave 
function can be estimated in the following way.
The higher partial waves are only substantial for large relative energies and 
are thus negligible in the FSI peak region.
In the QF peak, the relative $nn$ momentum is roughly 80~MeV/$c$, which means 
that the $S$-wave phase shift is $\delta_0\alt60^\circ$, while the $P$-wave 
phase shifts are typically $\delta_1\alt5^\circ$.
The $P$- to $S$-wave amplitude ratio can then be estimated as 
$A_1/A_0\sim \sin\delta_1/\sin\delta_0=0.10$.
\begin{figure}[t]
\includegraphics{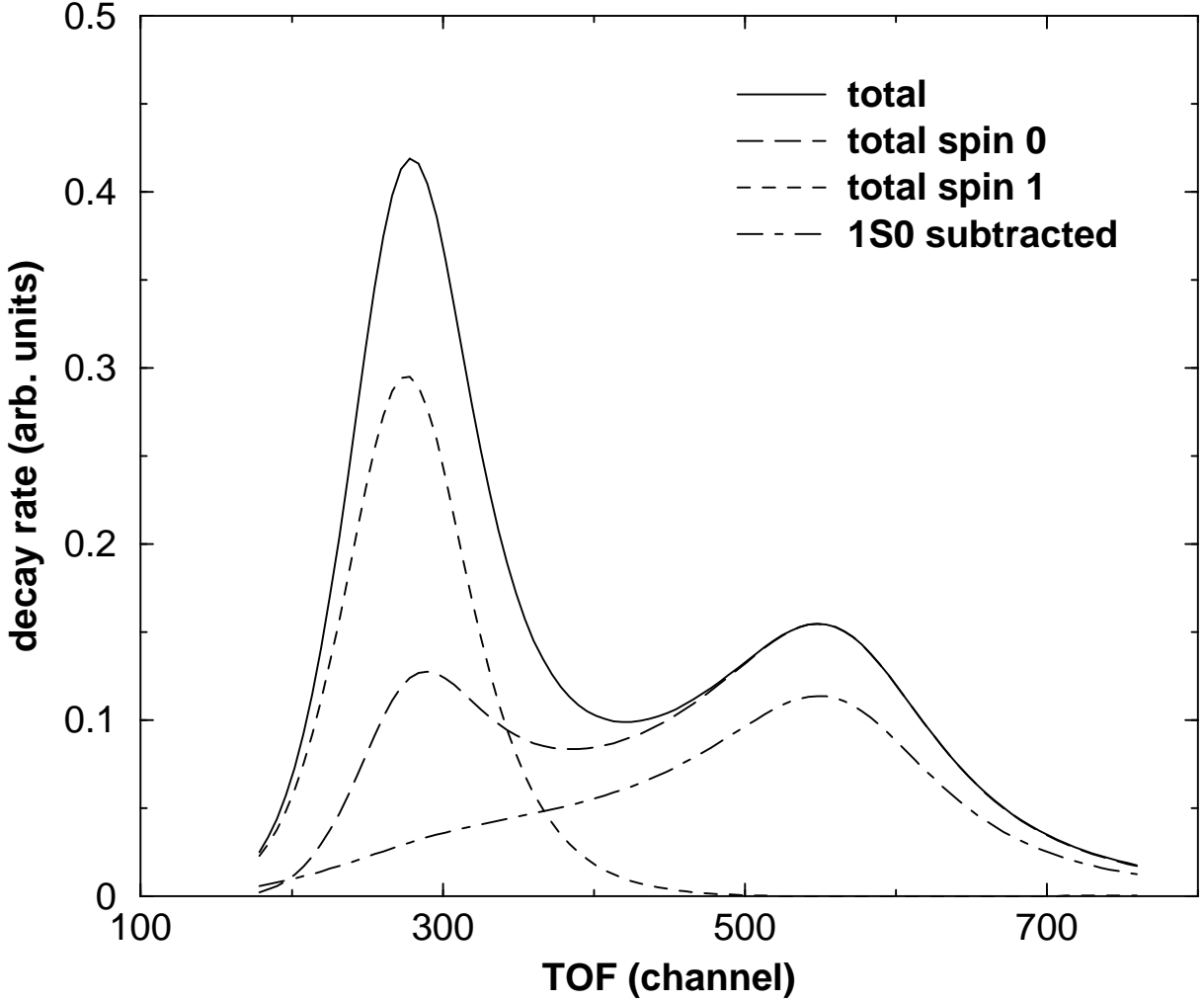}
\caption{The decay rate separated into spin-0 and spin-1 contributions.
The subtracted FSI $^1S_0$ partial wave is also distinguished.}
\label{fig:hipw}
\end{figure}
 From Fig.~\ref{fig:hipw} the $S$-wave FSI amplitude at the QF peak is
$A_0=\sqrt{0.030}=0.17$ and thus the $P$-wave FSI amplitude is
$A_1\sim0.10A_0=0.017$.  However, since the $P$-waves are spin-1 and
the $S$-waves spin-0 and the two do not interfere, the influence of
the $P$-wave should be related to the QF spin-1 amplitude, which is
$B_1=\sqrt{0.295}=0.543$.  The error in the calculated QF peak is then
$2|A_1||B_1||\cos\theta|$, where $\theta$ is the unknown phase angle
between $A_1$ and $B_1$.  Using the maximal possible error (setting
$\cos\theta=1$) seems overly pessimistic, so we instead choose the
average $\langle|\cos\theta|\rangle=2/\pi$.  The relative error at the
QF peak is then $\frac{4}{\pi}|A_1||B_1|/\Gamma_{\rm QF}\sim2.9\%$,
yielding an error in the extracted scattering length of 2.4\% or
0.43~fm.  This should be regarded as a conservative estimate of the
error of neglecting $P$-waves for two reasons.  Firstly, the $P$-waves
could interfere destructively with each other.  And secondly, we
implicitly assume that the radial integral for $P$-waves is of the
same magnitude as for $S$-waves, whereas it is probably smaller.
Most importantly, it is possible to actually calculate and include the
$P$-waves.  This error can thus easily be pushed to higher partial
waves and so made substantially smaller.  This will be done in future
work.

Note that this $P$-wave error is much larger than the one estimated by
GGS~\cite{GGS}.  The reason is that the $P$-waves only contribute at
large relative energy, \IE, under the QF peak, where they can
interfere with the QF amplitudes, thus changing the QF to FSI peak
ratio and the extracted $a_{nn}$.  The GGS error estimate assumes that
the $nn$ opening angle is smaller than $30^\circ$, which restricts the
kinematics to the FSI peak region only and thus does not apply to the
entire range of neutron energies used in the LAMPF extraction.  Also
in the work of de~T\'eramond~\EA~\cite{deTeramond} as used by the PSI
group~\cite{Gabioudetal}, only the FSI peak region is fitted, which
gives a small $P$-wave contribution.  Thus, as far as we can
ascertain, our analysis is the first that estimates
the interference of the FSI $P$-waves with the spin-1 QF amplitude.

If this effect was not included in the analysis of the data in
Ref.~\cite{LAMPF} then the $\Delta a_{nn}$ of approximately 0.43 fm we
have found here should be included in the theoretical uncertainty
quoted in that work. However, correspondence with one of the authors
of Ref.~\cite{LAMPF} suggests that FSI in $NN$ $P$-waves was included in
the version of the GGS model used for the
$a_{nn}$ extraction there~\cite{Ben}.  This source of uncertainty
would then not be present in Ref.~\cite{LAMPF}'s value for $a_{nn}$.

\subsubsection{Sensitivity to $r_0$}

An estimate of a change in $r_0$ due to CSB can be obtained by
assuming that the relative change in $r_0$ is similar to the relative
change in $a_{NN}$.  
Thus $\frac{\Delta r_0}{r_0}\approx\frac{\Delta a}{a}$ so that 
$\Delta r_0=\frac{r_0}{a}\Delta a=\frac{2.75}{18}1.5=0.23$~fm.  
The sensitivity to the effective-range parameter $r_0$ was tested by varying 
it away from its nominal value $2.75$~fm, using a conservative spread of 
$\pm 0.25$~fm.  
This changes the FSI peak by $1.4\%$ (after rescaling to the QF peak) and 
thus indicates a change in the extracted $a_{nn}$ of 1.2\% or 0.21~fm.
On the other hand, the error suggested by analysis of different
experimental determinations of $r_0$ is $\pm0.11$~fm~\cite{Slaus}. 
If $r_0$ is instead varied over this narrower range the resultant $\Delta
a_{nn}$ is only 0.5\% or 0.09~fm.
We will use the latter, smaller, error in our error budget.

\subsection{Error budget}

The errors are summarized in Table~\ref{tab:errors}.
\begin{table}[t]
\caption{Error budget for the extraction of $a_{nn}$ from the 
$\pi^-d\to nn\gamma$ reaction as it was performed in Ref.~\cite{LAMPF}.
The calculation of the absolute errors assumes a scattering length of -18~fm.
The total error is summed in quadrature.}
\label{tab:errors}
\begin{ruledtabular}
\begin{tabular}{rdd}
Source & \multicolumn{1}{c}{Relative error (\%)} & 
\multicolumn{1}{c}{Absolute error (fm)} \\
\hline
Off-shell         &  0.07  &  0.01 \\ 
Boost             & <0.11  & <0.02 \\ 
Subthreshold      &  0.95  &  0.17 \\ 
$O(Q^4)$ 2B       &  0.7   &  0.12 \\ 
$r_0$             &  0.5   &  0.09 \\ 
Dep.\ on $R_d$    &  0.55  &  0.10 \\ 
$p$-wave in FSI   & <2.4   & <0.43 \\ 
Dep.\ on $R_{nn}$ & <3.3   & <0.60 \\ 
total             & <4.3   & <0.78    
\end{tabular}
\end{ruledtabular}
\label{table:errors}
\end{table}
The first four errors are due to uncertainties in the amplitudes,
while the last four are due to the wave functions~\footnote{We realize
that such a separation is, strictly speaking, not meaningful, since
unitary transformations can be employed to trade ``wave-function''
effects for ``operator'' effects. However, the separation makes sense
within the approach to the calculation we have adopted here.}. 
We consider the total error of $<4.3\%$ to be a very conservative estimate.  

Note that if $a_{nn}$ is extracted only from data in the FSI region
then the last two errors drop to 0.2\% and 0.5\% respectively, while a
number of the other errors listed in Table~\ref{table:errors} are
also reduced.  We find that an extraction performed using only data from 
this section of the neutron time-of-flight spectrum would have a
theoretical uncertainty of $\pm 0.2$ fm. This confirms the conclusion
of GGS from thirty years ago. The significantly reduced theoretical
uncertainty comes at a price though: one must sacrifice the large
number of counts acquired under the QF peak. We have argued above that
the last two errors quoted in Table~\ref{table:errors} can be
decreased by additional theoretical work on radiative pion-capture on
deuterium, and therefore we hold out hope that in future a $\chi$EFT
extraction of $a_{nn}$ which has an accuracy of $\pm 0.3$~fm (or better) 
and uses the full neutron spectrum obtained in Ref.~\cite{LAMPF} can be
performed.

One reason for this optimism is the convergence of the chiral
expansion for this reaction, which can be made more explicit by computing the
QF to FSI peak ratio for the different orders.  From Fig.~\ref{fig:KR}
we obtain
\begin{equation}
  \frac{\Gamma_{\rm QF}}{\Gamma_{\rm FSI}}=(2.580+0.014+0.112\pm0.039)
  (1\pm0.05),
\end{equation}
where the first parenthesis contains (in order) the contribution of
the LO, NLO, NNLO, and the error in the chiral expansion.  The second
parenthesis shows the error due to effects in the wave functions.
Note that modifying the wave functions by including two-pion
exchanges, $P$-waves, or different short-distance dynamics would
already change the LO calculation, which is why we choose to write
this error as an overall factor.
The smallness of the NLO and NNLO one-body terms can perhaps be an effect of 
the particular kinematics of the present problem, especially that the pion 
momentum is vanishing.
On the other hand, the comparatively large NNLO two-body contribution is most 
likely a result of a combination of two effects:
Firstly, the two-body currents allow for momentum sharing between the nucleons,
which would be of importance in the QF region.
Secondly, in the leading two-body diagram [Fig.~\ref{fig:two}(a)], the 
coulombic propagator was power-counted as $1/\mu^2$.
However, because of the small deuteron binding energy, the typical momentum 
is instead of the order $\gamma=\sqrt{MB}=45.7$~MeV~\cite{pid}.
Since $\gamma\ll\mu$ this further enhances this diagram.

\section{Conclusions}
\label{sec:end}
In this paper we have calculated the $\pi^-d\to nn\gamma$ reaction,
using $\chi$PT pion-photon amplitudes and EFT-inspired wave functions.
The errors in the extracted scattering length from the operators are
of the order 1\%.  These errors include effects that were not
considered by Gibbs, Gibson, and Stephenson (GGS)~\cite{GGS}, \EG,
errors from extrapolating the single-nucleon amplitudes sub-threshold,
the boost of the $\gamma n\to\pi^-p$ amplitude from the $\gamma n$
rest frame to c.m., the effects of off-shell nucleons, and more
complicated two-body mechanisms.  A key improvement is that we have
included the full two-body amplitude at third chiral order and have
found that on the scale of the other errors it has a substantial
influence on the extraction of the scattering length.

Nevertheless, if $a_{nn}$ is extracted from the FSI region alone our
analysis within $\chi$EFT confirms GGS's result for the theoretical
uncertainties, putting them at $\pm 0.2$ fm.  On the other hand,
if---as was done in the most recent $a_{nn}$
extraction~\cite{LAMPF}---the entire shape of the neutron spectrum,
including both the QF and FSI peaks, is used for the extraction, then
the uncertainty in the scattering wave function at small distances and
the neglect of higher partial waves is a potentially large source of
errors, maybe as large as 4.3\%.  This might seem like a large
uncertainty, since it is almost three times larger than the 1.5\%
estimated by GGS.  But, as was argued in Sec.~\ref{sec:errorwf}, some
of the assumptions behind their error estimate do not seem to apply
for the entire kinematic range spanned by the data from the LAMPF
experiment.  This tempts us to suggest that the error estimate given
in Ref.~\cite{LAMPF} is optimistic and should be increased.

We plan to improve our model in the near future by constraining the
short-distance part of the $nn$ wave function using other observables
and by incorporating higher partial waves.  We will report these
results in a future publication. We also plan to fold our model with
the neutron detector acceptance and the experimental geometry in order
to extract the $nn$ scattering length from the data of
Ref.~\cite{LAMPF}.

Overall we conclude that the $\pi^-d\to nn\gamma$ reaction has some
very desirable features that makes it extremely suitable for the
extraction of the neutron-neutron scattering length.  The vanishing
pion momentum obviously favors a $\chi$PT calculation and also reduces
the number of contributing terms dramatically, leading to the
dominance of the Kroll-Ruderman term.  The fact that the extraction is
done by fitting the shape of time-of-flight spectra rather than an
absolute decay rate reduces many errors further still.

The reaction $\gamma d\to nn\pi^+$ could be used as an alternative and
complementary way to extract the neutron-neutron scattering length.
This reaction has been considered before, see the review \cite{Laget}
and later papers, \EG, \cite{Ulla}.
A chiral calculation should be feasible for this reaction, and could
benefit from the work of the present paper.
With a threshold photon laboratory energy of $149$~MeV it should be 
accessible at existing experimental facilities, \EG, HI$\gamma$S@TUNL after 
the planned upgrade and MAX-lab in Lund, Sweden.
After the submission of the manuscript, a calculation of $\gamma d\to nn\pi^+$
using chiral perturbation theory along lines similar to ours
has become available~\cite{Lensky:2005hb}.

\begin{acknowledgments}
We are grateful to T.~Hemmert for clarifications regarding the
single-nucleon radiative pion absorption amplitudes, and to
B.~F.~Gibson and C.~Howell for information on details of the
theoretical model used in Ref.~\cite{LAMPF}. A. G. thanks
C.~J.~Horowitz for discussions that led to a better understanding of
the scattering wave functions.  This work was supported by the DOE
grants DE-FG02-93ER40756 and DE-FG02-02ER41218.
\end{acknowledgments}

\bibliographystyle{apsrev}

\end{document}